\documentclass{iopart}
\usepackage[dvips]{graphicx}
\usepackage{amssymb}

\eqnobysec

\begin{document}

\title{The Fermi-Hubbard model at unitarity
}

\author{Evgeni Burovski$^1$, Nikolay Prokof'ev$^{1,2,3}$, Boris Svistunov$^{1,2}$ and Matthias Troyer$^{4}$ }

\address{$^1$ Department of Physics, University of Massachusetts,
Amherst, MA 01003, USA}%
\address{$^2$ Russian Research Centre ``Kurchatov Institute'',
123182 Moscow, Russia}%
\address{$^3$ BEC-INFM, Dipartimento di
Fisica, Universita di Trento, Via Sommarive 14, I-38050 Povo,
Italy}%
\address{$^4$ Theoretische Physik, ETH Z{\"urich},
CH-8093 Z{\"u}rich, Switzerland}

\ead{burovski@physics.umass.edu}

\begin{abstract}
We simulate the dilute attractive Fermi-Hubbard model
in the unitarity regime using a diagrammatic determinant Monte Carlo algorithm with
worm-type updates.
We obtain the
dependence of the critical temperature on the filling factor $\nu$ and, by
extrapolating to $\nu \to 0$, determine the universal critical temperature
of the continuum unitary Fermi gas in units of Fermi energy:
$T_c/\varepsilon_F=0.152(7)$. We also determine the thermodynamic functions
and show how the Monte Carlo results can be used for accurate
thermometry of a trapped unitary gas.
\end{abstract}


\maketitle

\section{Introduction}
\label{sec:intro}

In recent years, ultracold atomic systems have served as a
controlled and tunable toolbox for studying many-body quantum
phenomena. The continuous tunability of the interaction by
Feshbach resonances makes these systems ideal candidates to study
the crossover from momentum space pairing in the
Bardeen--Cooper--Schrieffer (BCS) theory to Bose-Einstein
condensate (BEC) of fermions bound into molecules. This BCS-BEC
crossover has been one of the most studied problems in recent
experiments in both magnetic or optical traps
\cite{BEC-BCS-expt,Hulet-PRL2005,Grimm-in-situ,Hulet-Science2006,expt-univ-energy,expt-ramp}
and optical lattices \cite{opt-latt-expt}.

Tuning across the Feshbach resonance, one traverses the whole
range of the gas parameter $k_F a_s$, where $k_F$ is the Fermi
momentum and $a_s$ is the s-wave scattering length. The regime of
$k_F |a_s| \ll 1$ (negative $a_s$) is described by
Bardeen--Cooper--Schrieffer (BCS) theory. At  $k_F a_s \ll 1$
(positive $a_s$) fermions pair into bosonic molecules and form a
Bose-Einstein condensate (BEC). At the microscopic scale, the BCS
and BEC regimes are radically different; however, the macroscopic,
and, in particular, critical behaviour is supposed to be
qualitatively the same for the whole range of $k_F a_s$: the
system undergoes the superfluid (SF) phase transition at a certain
critical temperature. Separating the BCS and BEC extremes is the
so-called unitarity point $(k_F a_s)^{-1} \to 0$. It is worth
noting that the unitarity regime is approximately realized in the
inner crust of the neutron stars, where the neutron-neutron
scattering length is nearly an order of magnitude larger than the
mean interparticle separation \cite{neutron-star}.

The Fermi gas at unitarity is a peculiar case of a strongly
interacting system with no interaction-related energy scale: the
divergent scattering length and any related energy scale drop out
completely. This gives rise to universality of the dilute gas
properties, in the sense that the only relevant energy scale left
in the system is given by the density, $n$. Because of this
universality one obtains a unified description of such diverse
systems as  cold atoms in magnetic or optical traps, Fermi-Hubbard
model in optical lattices and inner crusts of neutron stars.

The theoretical description of the Fermi gas in the  BCS-BEC
crossover regime is a major challenge, since the system features
no small parameter on which one could build a theory in a rigorous
way. The original analytical treatments were confined to zero
temperature and were based on the extension of the BCS-type
many-body wave function \cite{classiki}. Most of the subsequent
elaborations are also of mean-field type (with or without remedies
for the effects of fluctuations)
\cite{NSR,Haussmann94,Randeria95,Holland-Timmermans,Ohashi-Griffin,Perali,LiuHu}. The
accuracy and reliability of such approximations is questionable
since they inevitably involve an uncontrollable approximation.

Numerical investigations of the unitary Fermi gases are
hampered by the sign problem, inherent to any Monte Carlo (MC)
simulations of fermion systems \cite{BinderLandau-book,TroyerWiese}. One way
of avoiding the sign problem at the expense of a systematic error
is the fixed-node Monte Carlo framework, which has been used to
study the ground state \cite{Giorgini-Carlson}.
The systematic error of the fixed-node Monte Carlo depends on
the quality of the variational ansatz for the nodal structure
of the many-body wave function and is not known precisely.
Only in a few exceptional cases can the sign problem  be avoided
without incurring systematic errors. One of such cases is given by
fermions with attractive contact interaction, for which a
number of sign-problem-free schemes has been introduced
\cite{Hirsch,Rombouts,Rubtsov,Kaplan}. Fortunately, this system
can be tuned to the unitarity regime. Still, despite a number of
calculations at finite temperature
\cite{Bulgac,Wingate,Lee-Schaefer}, an accurate description of
the finite-temperature properties of the unitary Fermi gas is
missing.

In the present paper, we simulate the Fermi-Hubbard model in the
unitary regime by means of a determinant diagrammatic Monte Carlo
method. By studying the dilute limit of the model, we extract
properties of the homogeneous continuum Fermi gas. A brief summary
of the main results has been given in Ref.~\cite{we-short}. Here we
provide a detailed description of the Monte Carlo scheme and
methods of data analysis. We also report new results relevant
to experimental realizations of the Fermi-Hubbard model in optical
lattices and trapped Fermi gases.

The Fermi-Hubbard model is defined by the Hamiltonian
\numparts
\begin{equation}
H = H_0 + H_1,  \\
\label{AHM:1}
\end{equation}
\begin{equation}
H_0 = \sum_{\mathbf{k}\sigma} \left( \epsilon_\mathbf{k} - \mu
\right) c^{\dagger}_{\mathbf{k}\sigma}c_{\mathbf{k}\sigma},\label{AHM:2}\\
\end{equation}
\begin{equation}
H_1 = U \sum_\mathbf{x} n_{\mathbf{x}\uparrow}
n_{\mathbf{x}\downarrow}\; . \label{AHM:3}
\end{equation}
\endnumparts
%
Here $c^{\dagger}_{\mathbf{k}\sigma}$ is a fermion creation
operator, $n_{\mathbf{x}\sigma} = c^{\dagger}_{\mathbf{x}\sigma}
c_{\mathbf{x}\sigma}$, $\sigma = \uparrow, \downarrow$ is the spin
index, $\mathbf{x}$ enumerates $L^3$ sites of the
three-dimensional (3D) simple cubic lattice with periodic boundary
conditions, the quasimomentum $\mathbf{k}$ spans the corresponding
Brillouin zone, $\epsilon_\mathbf{k}=2t \sum_{\alpha=1}^{3}(1-\cos
k_\alpha a)$ is the single-particle tight-binding spectrum, $a$
and $t$ are the lattice spacing and the hopping amplitude,
respectively, $\mu$ stands for the chemical potential and $U<0$ is
the on-site attraction. Without loss of generality we henceforth
set $a$ and $t$ equal to unity;  the effective mass at the bottom
of the band is then $m=1/2$.

In Sec.\ \ref{sec:2body} we will study the two-body problem at
zero temperature, show how the Hubbard model can be used to study
the continuum unitary gas and investigate the functional structure
of lattice corrections to the continuum behaviour. In Sec.\
\ref{sec:ddmc} we discuss the finite-temperature diagrammatic
expansion for the Hubbard model (Sec. \ref{ssec:matsubara}), and
give a qualitative description of the Monte Carlo procedure to sum
the diagrammatic series (Sec.\ref{ssec:MC}), with details of the
updating procedures given in Appendix \ref{sec:updates}.  In order
to extract the thermodynamic limit properties from MC data, we use
finite-size scaling analysis described in  Sec.\ \ref{sec:sense}.
Section\ \ref{sec:thermodynamics} gives an overview of the scaling
functions describing thermodynamics of the unitary gas, and
results are presented and discussed in Sec.\ \ref{sec:results}.

\section{Two-body problem}
\label{sec:2body}

\begin{figure}
\includegraphics[width=0.75\columnwidth,keepaspectratio=true]{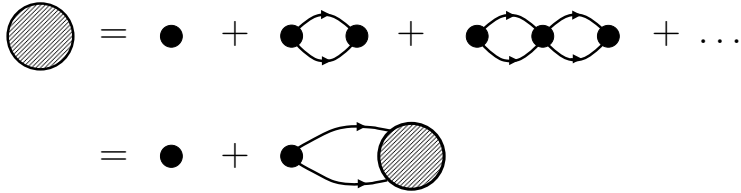}
\caption{ Diagrammatic series for the vertex insertion
$\Gamma(\xi,\mathbf{p})$ (heavy dot). Small dots represent the
bare Hubbard interaction $U$, and lines are the single-particle
propagators. }
\label{fig:ladder}
\end{figure}

Consider a quantum mechanical problem of two  fermions at zero
temperature described by the Hamiltonian
(\ref{AHM:1})--(\ref{AHM:3}) with $\mu=0$. The most
straightforward way to tackle this problem is within the
diagrammatic technique in the momentum-frequency representation
\cite{Fetter-Walecka,IXtom}, which, in the present case, is built
on four-point vertices, $U$, with two incoming (spin-$\uparrow$
and spin-$\downarrow$) and two outgoing (spin-$\uparrow$ and
spin-$\downarrow$) ends, connected by single-particle propagators.
The scattering of two particles is then described by a series of
ladder diagrams \cite[\S16]{IXtom} shown in \fref{fig:ladder}.
Ladder diagrams can be summed by introducing the vertex insertion
$\Gamma(\xi,\mathbf{p})$, which depends on frequency $\xi$ and
momentum $\mathbf{p}$. Since $\Gamma (0,0)$ is proportional to the
scattering amplitude, the unitarity limit corresponds to
$\Gamma(\xi\to 0,\mathbf{p}\to \mathbf{0})\to \infty$. The
summation depicted in \fref{fig:ladder} leads to
\begin{equation}
\Gamma^{-1}(\xi,\mathbf{p}) = U^{-1} + \Pi(\xi,\mathbf{p})\; ,
\label{Dyson}
\end{equation}
where $\Pi(\xi,\mathbf{p})$ is the polarization operator (the
integration is over the Brillouin zone):
\begin{equation}
\Pi(\xi,\mathbf{p}) = \int_{\mathrm{BZ}}
\frac{\rmd\mathbf{k}}{(2\pi)^3}
\frac{1}{\xi+\epsilon_{\mathbf{p}/2+\mathbf{k}}+\epsilon_{\mathbf{p}/2-\mathbf{k}}}
\; .
\label{polariz}
\end{equation}

It immediately follows from Eq.\ (\ref{Dyson}) and (\ref{polariz})
that the unitary limit corresponds to $U=U_*$, where
\begin{equation}
U^{-1}_* = -\Pi(0,\mathbf{0}) = -\int_\mathrm{BZ}
\frac{\rmd\mathbf{k}}{(2\pi)^3} \,
\frac{1}{2\epsilon_\mathbf{k}}\; .
\label{U-unitary}
\end{equation}
A straightforward numeric integration yields $U_*\approx -7.915t$.

In the limit of vanishing filling factor $\nu \to 0$ for the
many-body problem of (\ref{AHM:1})--(\ref{AHM:3})  the typical
values of $\xi $ and $p$ are related to the Fermi energy $\xi \sim
\epsilon_F \sim \nu^{2/3}$ and Fermi momentum $p \sim k_F \sim
\nu^{2/3}$ and are small compared to the bandwidth and reciprocal
lattice vector, respectively. In zero-th order with respect to
$\nu$, the lattice system is identical to the continuum one.
Indeed, by combining (\ref{Dyson}), (\ref{polariz}) and
(\ref{U-unitary}), we get
\begin{equation}
\Gamma^{-1}(\xi,\mathbf{p}) = \int_{\mathrm{BZ}}
\frac{\rmd\mathbf{k}}{(2\pi)^3} \left[
\frac{1}{\xi+\epsilon_{\mathbf{p}/2+\mathbf{k}}+
\epsilon_{\mathbf{p}/2-\mathbf{k}}} -
\frac{1}{2\epsilon_{\mathbf{k}}} \right]\; ,
\label{Gamma-latt}
\end{equation}
and observe that for small $\xi $ and $p$ one can replace
$\varepsilon({\bf k})$ with $k^2$ and extend integration over
$\rmd\mathbf{k}$ to the whole momentum space with the result
\begin{equation}
\Gamma^{-1}_\mathrm{cont}(\xi,\mathbf{p}) =
-\frac{m^{3/2}}{4\pi}\sqrt{\xi+\frac{p^2}{4m}} \; .
\label{Gamma-cont}
\end{equation}
With this form of $\Gamma_\mathrm{cont}$, and particle propagators
based on the parabolic dispersion relation one recovers the
continuum limit behaviour.

Now we are in a position to treat the lattice corrections. The
first correction should come from $\Gamma$, not from propagators,
since only in $\Gamma$ large momenta play a special role
due to resonance in the two-particle channel.
In the lowest non-vanishing order in $\xi$ and $\mathbf{p}$ we
have (the summation over repeating subscripts is implied)
\begin{equation}
\Gamma^{-1} \approx - \int_\mathrm{BZ} {\rmd{\bf k}\over (2\pi)^3}
\, \frac{\xi + (1/4)(\partial^2 \varepsilon_\mathbf{k}/\partial
k_i
\partial k_j) p_i p_j }
{4\varepsilon_\mathbf{k}^2}  \; .
\end{equation}
with the difference between the lattice and continuous model given
by
\begin{equation}
\Gamma^{-1} - \Gamma^{-1}_\mathrm{cont}\; \approx\;  \frac{\xi}{4}
A + \frac{p^2}{16} B\; ,
\end{equation}
where
%
\begin{eqnarray} A = \int {\rmd{\bf k}\over (2\pi)^3} \,  {1 \over (k^2/2m)^2} -
\int_\mathrm{BZ} {\rmd{\bf k}\over (2\pi)^3} \, {1\over
\varepsilon_\mathbf{k}^2} \; ,\label{Gamma-A}\\
B = \int {\rmd{\bf k}\over (2\pi)^3} \,  {1/m \over (k^2/2m)^2} -
\int_\mathrm{BZ} {\rmd{\bf k}\over (2\pi)^3} \, {(\partial^2
\varepsilon_\mathbf{k}/\partial k_x \partial k_x)\over
\varepsilon_\mathbf{k}^2}\;  . \label{Gamma-B}
\end{eqnarray}


In the limit of $\xi\to 0$ and $p\to 0$, we have
$\Gamma^{-1} \approx \Gamma^{-1}_\mathrm{cont} \sim k_F \sim
\nu^{1/3}$, and
$\Gamma^{-1}-\Gamma^{-1}_\mathrm{cont} \sim k_F^2 \sim \nu^{2/3} $.
Hence, the leading lattice correction is of the form
\begin{equation}
\frac{\Gamma(\xi,\mathbf{p}) -
\Gamma_\mathrm{cont}(\xi,\mathbf{p})}{\Gamma(\xi,\mathbf{p})} \;
\sim\;  \nu^{1/3}\; .
\label{nu-13}
\end{equation}
Incidentally, Eqs. (\ref{Gamma-A}) and (\ref{Gamma-B}) hint into
an intriguing possibility of completely suppressing the
leading-order lattice correction by tuning the single-particle
spectrum $\epsilon_\mathbf{k}$ so that $A=B=0$.  We did not
explore this possibility in the present study.

\section{Determinant Diagrammatic Monte Carlo}
\label{sec:ddmc}

The diagrammatic technique employed in the previous section is not
particularly convenient for numerical studies. In this
section, we briefly review the Matsubara technique and then
present a Monte Carlo scheme of summing the resultant
diagrammatic series.
\subsection{Rubtsov's representation}
\label{ssec:matsubara} %

To construct a diagrammatic expansion for the model
(\ref{AHM:1})--(\ref{AHM:3}) we follow Refs.~\cite{Rubtsov,wePRB}
and consider the statistical operator in the
coordinate---imaginary time representation. In the interaction
picture we get:
\begin{equation}
\exp(-\beta H)\;  =\;  \exp(-\beta H_0)\,  \mathcal{T}_\tau
\exp\left( - \int_0^\beta \, \rmd\tau H_1(\tau)\right)\; ,
\label{statOper}
\end{equation}
where $\beta$ is an inverse temperature, $H_1(\tau) = e^{\tau H_0}
H_1 e^{-\tau H_0} $, and $\mathcal{T}_\tau$ stands for the
imaginary time ordering.

\begin{figure}
\includegraphics[width=0.75\columnwidth,keepaspectratio=true]{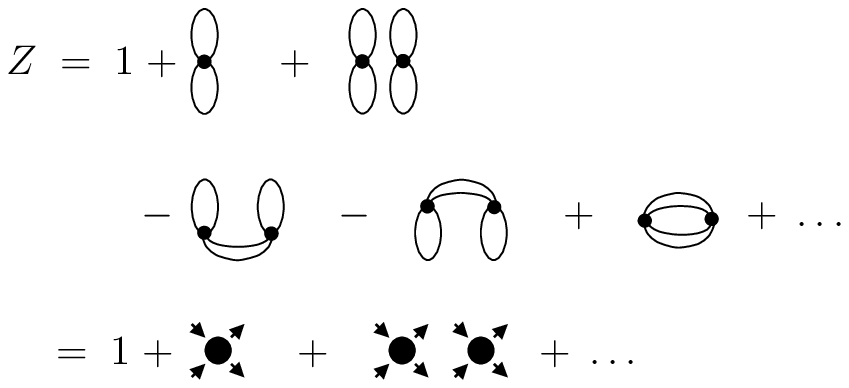}
\caption{ Diagrammatic series for the partition function. Upper
line is the graphical representation of the series
(\ref{Z-diagr}), lower line depicts Eq.\ (\ref{Z-summed}). Diagram
signs are shown explicitly.}
\label{fig:Z}
\end{figure}

Expanding Eq.\ (\ref{statOper}) in powers of $H_1$, one obtains
for the partition function:
%
\begin{equation}
\fl
Z \; =\;  \sum_{n=0}^\infty (-U)^n \sum_{\mathbf{x}_1 \dots \mathbf{x}_n}%
\int_{0<\tau_1<\tau_2< \dots < \beta} \prod_{j=1}^{n} \rmd\tau_j %
\Tr\left[ e^{-\beta H_0} \prod_{j=1}^{n}%
c_\uparrow^\dagger(\mathbf{x}_j \tau_j) c_\uparrow(\mathbf{x}_j \tau_j)%
c_\downarrow^\dagger(\mathbf{x}_j \tau_j) c_\downarrow(\mathbf{x}_j \tau_j)%
 \right]\; .
\label{Z-diagr}
\end{equation}
%
Expansion (\ref{Z-diagr}) generates the standard set of Feynman
diagrams. Graphically, the diagrams are similar to those of Sec.\
\ref{sec:2body}, and consist of the four-point vertices, $U$,
connected by the single-particle propagators $G_\sigma^{(0)}
(\mathbf{x}_i-\mathbf{x}_j, \tau_i-\tau_j; \mu,\beta) = - \Tr
\left[\mathcal{T}_\tau e^{-\beta H_0}
c_\sigma^\dagger(\mathbf{x}_i \tau_i)c_\sigma(\mathbf{x}_j \tau_j)%
\right]$. The $p$-th order of the perturbation theory is then
graphically given by a set of $(p!)^2$ possible interconnections
of vertices by propagators, see \fref{fig:Z}.

The diagrammatic expansion (\ref{Z-diagr}) is \emph{unsuitable}
for the direct Monte Carlo simulation since it has a sign problem
built in: different terms in the series have different signs --- a
closed fermion loop brings in an extra minus sign
\cite{Fetter-Walecka}. The trick is to consider all
diagrams of a given order $p$ with the fixed vertex configuration
\begin{equation}
\mathcal{S}_p = \{ (\mathbf{x}_j,\tau_j),~~j=1,\dots,p  \}
\label{conf-p}
\end{equation}
as one.  This implies summation over the $(p!)^2$ ways of
connecting vertices by propagators. Upon summation, Eq.\
(\ref{Z-diagr}) takes on the form \cite{Rubtsov}:
\begin{equation}
Z = \sum_{p=0}^\infty (-U)^p \sum_{\mathcal{S}_p} \det
\mathbf{A}^{\uparrow}(\mathcal{S}_p) \det
\mathbf{A}^{\downarrow}(\mathcal{S}_p)\; ,
\label{Z-summed}
\end{equation}
where
\begin{equation}
\sum_{\mathcal{S}_p} \equiv \sum_{\mathbf{x}_1 \dots \mathbf{x}_p}%
\int_{0<\tau_1<\tau_2< \dots < \tau_p < \beta} \prod_{j=1}^{p}
\rmd\tau_j\; ,
\label{S}
\end{equation}
and $\mathbf{A}^{\sigma}(\mathcal{S}_p)$ are the $p \times p$
matrices built on the single-particle propagators:
\begin{equation}
A^{\sigma}_{ij}(\mathcal{S}_p) = G^{(0)}_\sigma (\mathbf{x}_i -
\mathbf{x}_j, \tau_i-\tau_j)\; ,~~~i,j=1,\dots,p\; .
\label{matrix}
\end{equation}

For equal number of spin-up and spin-down particles $\det
\mathbf{A}^\uparrow \det \mathbf{A}^\downarrow = | \det \mathbf{A}
|^2$, and \emph{the sign problem is absent}. \footnote{At half
filling, the sign of $U$ changes if the hole representation is
used for one of the spin components. Hence, this method is also
applicable to the half-filled repulsive Hubbard model. }
Graphically, Feynman diagrams in this representation are just
collections of vertices, see \fref{fig:Z}. For future use, we
define the set of all possible vertex configurations
(\ref{conf-p}) by $\mathfrak{S}^{(Z)}$, \textit{i.e.,}
$\mathfrak{S}^{(Z)} = \{p, \{ \mathcal{S}_p \} \}$.

The following two-point pair correlation function will prove useful:
\begin{equation}
G_2(\mathbf{x}\tau; \mathbf{x}'\tau')\; =\;  %
\left\langle\,  \mathcal{T}_\tau P(\mathbf{x},\tau)
P^\dagger(\mathbf{x}',\tau')\,  \right\rangle \; \equiv \;
{g_2(\mathbf{x}\tau; \mathbf{x}'\tau') \over Z}\; ,
\label{corr}
\end{equation}
\begin{equation}
g_2(\mathbf{x}\tau; \mathbf{x}'\tau') \; =\; {\rm Tr}\,
\mathcal{T}_\tau P(\mathbf{x},\tau) P^\dagger(\mathbf{x}',\tau')\,
{\rm e}^{-\beta H} \; ,
\label{g_2}
\end{equation}
where $P(\mathbf{x},\tau)$ and $P^\dagger(\mathbf{x}',\tau')$ are
the pair annihilation and creation operators in the Heisenberg
picture, respectively: $P(\mathbf{x},\tau) =
c_{\mathbf{x}\uparrow}(\tau) c_{\mathbf{x}\downarrow}(\tau)$. The
non-zero asymptotic value of $G_2(\mathbf{x}\tau;
\mathbf{x}'\tau')$ as $|\mathbf{x}-\mathbf{x}'| \to \infty$ is
proportional to the condensate density.

\begin{figure}
\includegraphics[width=0.75\columnwidth,keepaspectratio=true]{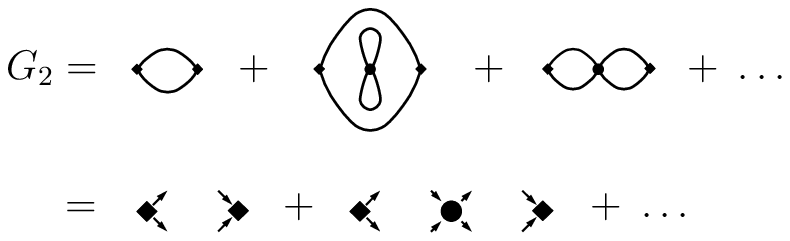}
\caption{ Diagrammatic series for the correlation function
(\ref{corr}). Diamonds represent the two-point
creation/annihilation operators $P$ and $P^\dagger$. }
\label{fig:G}
\end{figure}

Feynman diagrams for $g_2(\mathbf{x}\tau; \mathbf{x}'\tau')$ are
similar to those for $Z$, but contain two extra elements: a pair
of  two-point vertices with two incoming (outgoing) ends which
represent $P(\mathbf{x},\tau)$ ( $P^\dagger(\mathbf{x}',\tau')$ ),
see \fref{fig:G}. The vertex configurations for the correlation
function~(\ref{corr}) slightly differ from those for the partition
function (\ref{conf-p}) by the presence of the two extra elements:
the configuration space for Eq.\ (\ref{corr}) is
$\mathfrak{S}^{(G)} = \{ p, \{ \tilde{\mathcal{S}}_{p} \} \}$,
with
\begin{equation}
\tilde{\mathcal{S}}_p = \{ P(\mathbf{x},\tau),\,
P^\dagger(\mathbf{x}',\tau'),\,
(\mathbf{x}_j,\tau_j),~~j=1,\dots,p \} \; .
\label{conf-p-corr}
\end{equation}
The partially summed diagrammatic expansion for
$g_2(\mathbf{x}\tau; \mathbf{x}'\tau')$ is similar to Eq.\
(\ref{Z-summed}):
\begin{equation}
g_2(\mathbf{x}\tau; \mathbf{x}'\tau') = \sum_{p=0}^\infty (-U)^p
\sum_{\tilde{\mathcal{S}}_p} \det
\widetilde{\mathbf{A}}^{\uparrow}(\tilde{\mathcal{S}}_p) \det
\widetilde{\mathbf{A}}^{\downarrow}(\tilde{\mathcal{S}}_p)\; ,
\label{G-summed}
\end{equation}
where $\widetilde{\mathbf{A}}^\sigma(\tilde{\mathcal{S}}_p)$ is a
$(p+1)\times(p+1)$ matrix which differs from Eq.\ (\ref{matrix})
only by that it has an extra
row $i_0$ and an extra column $j_0$ such that
$\widetilde{A}^\sigma_{ij_0} = G^{(0)}_\sigma (\mathbf{x}_i -
\mathbf{x}, \tau_i-\tau)$ and $\widetilde{A}^\sigma_{i_0 j} =
G^{(0)}_\sigma (\mathbf{x}' - \mathbf{x}_j, \tau' - \tau_j)$.

Below we deal only with equal number of spin-up and spin-down
fermions and (for the sake of brevity) suppress the spin indices
wherever possible. We also peruse the generic notation (with
superscripts) $\mathcal{D}(\mathcal{S}_p)$ for $p$-th order terms
of the diagrammatic expansions similar to (\ref{Z-summed}), e.g.,
$\mathcal{D}^{(Z)}(\mathcal{S}_p) = (-U)^{p}\bigl| \det
\mathbf{A}(\mathcal{S}_p) \bigr|^2$. To simplify the notation we
also omit superscripts if this does not lead to ambiguity.

\subsection{Diagrammatic Monte Carlo and Worm algorithm} \label{ssec:MC}

Equations\ (\ref{Z-summed}) and (\ref{G-summed}) have similar general
structure of a series of integrals and sums with ever increasing
number of integration variables and summations. In Refs.\ %
\cite{diagrMC,polaron00} it has been shown how to arrange a
numerical procedure that sums such convergent series. To this end
one considers the space of all possible vertex configurations
$\mathfrak{S}$ (~for the series (\ref{Z-summed}) $\mathfrak{S}
\equiv \mathfrak{S}^{(Z)}$, while for the series (\ref{G-summed})
$\mathfrak{S} \equiv \mathfrak{S}^{(G)}$~), with the ``weight''
$\mathcal{D}(\mathcal{S}_p)$ associated with each element of the
space. One then uses the Metropolis principle \cite{Metropolis} to
arrange a stochastic Markov process which sequentially generates
vertex configurations $\mathcal{S}_p$ according to their weights
$\mathcal{D}(\mathcal{S}_p)$, thus sampling the space
$\mathfrak{S}$. In the course of sampling, one also collects
statistics for observables in the form of MC estimators, see Sec.\
\ref{ssec:estimators}.

The stochastic process consists of elementary MC updates performed
on vertex configurations $\mathcal{S}_p$. The set of updates is
problem-specific being restricted only by the requirements of (i)
the ergodicity, \textit{i.e.,} given a particular diagram
$\mathcal{S}_p$ it takes a finite number of steps to convert it
into any other diagram $\mathcal{S}'_q$, and (ii) detailed
balance, \textit{i.e.,} relative contributions of diagrams
$\mathcal{S}_p$ and $\mathcal{S}'_q$ to the statistics is given by
the ratio of their weights,
$\mathcal{D}(\mathcal{S}_p)/\mathcal{D}(\mathcal{S}'_q)$. The set
of updates satisfying these requirements is not unique, the
freedom is used to maximize the efficiency of simulations, as
explained in detail in  \ref{sec:updates}.

In view of close similarity between the diagrammatic expansions
(\ref{Z-summed}) and (\ref{G-summed}) it is advantageous to
construct a Monte Carlo process which samples these two series in
a single simulation. This way one has access to both diagonal,
\textit{e.g.,} energy, and off-diagonal properties, \textit{e.g.,}
the superfluid response. An efficient way of performing such
concurrent simulation is provided by the worm algorithm, which was
originally devised for the worldline Monte Carlo simulations
\cite{worm}. In the context of the diagrammatic determinant Monte
Carlo, the generic worm algorithm principles imply the following.
First, one works in the joint configuration space
$\mathfrak{S}^{(Z)} \cup \mathfrak{S}^{(G)} $, accommodating
diagrams (\ref{conf-p}) and (\ref{conf-p-corr}). Second, all the
updates are performed {\it exclusively} in terms of the two-point
vertices $P(\mathbf{x},\tau)$ and
$P^\dagger(\mathbf{x},\tau)$---through their
creation/annihilation, motion, and ``interactions" with adjacent
vertices. The worm-type updating procedures are further detailed
in  \ref{sec:updates}. Within the worm-algorithm framework, the
configuration spaces $\mathfrak{S}^{(Z)}$ and $\mathfrak{S}^{(G)}$
are disjoint subsets of one extended configuration space. In what
follows we will refer to them as $Z$-(or ``diagonal'') and $G$-(or
``off-diagonal'') sectors of the configuration space.

Formally, the extended configuration space corresponds to the
generalized partition function
\begin{equation}
Z_\mathrm{W}\;  =\;  Z + \zeta \sum_{\mathbf{x}, \mathbf{x}'}
\int_0^{\beta} \rmd\tau \int_0^{\beta}\rmd\tau'
g_2(\mathbf{x}\tau;\mathbf{x}'\tau')\; ,
\label{Z-worm}
\end{equation}
where the value of $\zeta$ is arbitrary~--- it controls the
relative statistics of $Z$- and $G$-sectors and the
efficiency of simulation.

\subsection{Monte Carlo estimators } \label{ssec:estimators}

Suppose we have an observable $X(\alpha)$, which depends on a set
of variables $\alpha$, \textit{e.g.} temperature and chemical
potential. A MC estimator for the observable $X$ is an expression
which, upon averaging over the sequence of MC
configurations converges to the expectation value of $X(\alpha)$.

In accordance with Eq.\ (\ref{Z-worm}), the simplest worm-algorithm
MC estimators are

\begin{equation}
\delta^{(Z)}(\mathcal{S}_p)\;  =\;  %
\cases{
1,& $\mathcal{S}_p \in \mathfrak{S}^{(Z)}$\; , \\
0,& $\widetilde{\mathcal{S}}_p \in \mathfrak{S}^{(G)}$\; ,\\
}
\label{delta-Z}
\end{equation}
and
\begin{equation}
\delta^{(G)}(\mathcal{S}_p)\;  =\;  %
\cases{
0,& $\mathcal{S}_p \in \mathfrak{S}^{(Z)}$\; , \\%
1,& $\widetilde{\mathcal{S}}_p \in \mathfrak{S}^{(G)}$\; .\\
}
 \label{delta-G}
\end{equation}

Their MC averages are
\begin{equation}
\bigl\langle \delta^{(Z)} \bigr\rangle_\mathrm{MC}\;
\longrightarrow \;  Z/Z_W\; , \label{delta-Z-MC}
\end{equation}
and
\begin{equation}
\bigl\langle \delta^{(G)} \bigr\rangle_\mathrm{MC}\;
\longrightarrow \; Z_W^{-1}\zeta \sum_{\mathbf{x}, \mathbf{x}'}
\int_0^{\beta} \rmd\tau \int_0^{\beta}\rmd\tau' \,
g_2(\mathbf{x}\tau;\mathbf{x}'\tau') \; , \label{delta-ZG-MC}
\end{equation}
where $\langle \dots \rangle_\mathrm{MC}$ denotes averaging over
the set of stochastically generated configurations. In
particular, for our purposes it will be quite useful that
\begin{equation}
\frac{\bigl\langle \delta^{(G)} \bigr\rangle_\mathrm{MC}}
{\bigl\langle \delta^{(Z)} \bigr\rangle_\mathrm{MC}}\;
\longrightarrow \; \zeta \sum_{\mathbf{x}, \mathbf{x}'}
\int_0^{\beta} \rmd\tau \int_0^{\beta}\rmd\tau' \,
G_2(\mathbf{x}\tau;\mathbf{x}'\tau') \; . \label{useful}
\end{equation}

The general rules for constructing an estimator for
a quantity $X(\alpha)$ specified by the diagrammatic expansion
\begin{equation}
X(\alpha) = \sum_{\mathcal{S}_p} \mathcal{D}^{(X)}(\alpha ; \mathcal{S}_p)\; ,
\label{XY}
\end{equation}
are standard. We adopt a convenient convention: If the actual
summation in (\ref{XY}) involves only a subset $\mathfrak{S}_0$ of
vertex configurations---a typical example is an expansion defined
within the $Z$-sector only---then we extend summation over the
entire configuration space by simply defining
$\mathcal{D}^{(X)}(\mathcal{S}_p \not\in \mathfrak{S}_0)\equiv 0$.
If vertex configurations $\mathcal{S}_p$ are sampled from the
probability density $\mathcal{D}^{(Z_W)}(\mathcal{S}_p)$ which
comes from the expansion for the generalized partition function:
\begin{equation}
Z_W(\alpha) = \sum_{\mathcal{S}_p} \mathcal{D}^{(Z_W)}(\alpha;
\mathcal{S}_p)\; , \label{Z_W_series}
\end{equation}
then the MC estimator for $X(\alpha)$ is derived from
\begin{equation}
\bigl\langle X \bigr\rangle \,\equiv \,
\frac{\bigl\langle \mathcal{Q}^{(X)}
\bigr\rangle_\mathrm{MC}}{\bigl\langle \delta^{(Z)}
\bigr\rangle_\mathrm{MC}} \; ,
\label{esti}
\end{equation}
as
\begin{equation}
\mathcal{Q}^{(X)}(\alpha;\mathcal{S}_p) \; =\;
\frac{\mathcal{D}^{(X)}(\alpha;\mathcal{S}_p)}
{\mathcal{D}^{(Z_W)}(\alpha;\mathcal{S}_p )} \; .
\label{esti-2}
\end{equation}
In what follows, by the estimator for a quantity
$x(\alpha)=X(\alpha)/Z(\alpha)$ we basically mean corresponding
function $\mathcal{Q}^{(X)}(\alpha;\mathcal{S}_p)$.

\subsubsection{Estimators for number density and kinetic energy}
\label{sssec:dens}

We start with the estimator for the number density. The
expectation value of the number density reads
\begin{equation}
\nu \; =\;  \frac{2 \Tr  c^{\dagger}_{\mathbf{x} \sigma}(\tau)
c_{\mathbf{x} \sigma}(\tau) e^{-\beta H} }{Z}\; .
\label{dens-gibbs}
\end{equation}
Here $(\mathbf{x},\tau)$ is an arbitrary space-time point (the
system is space/time translational invariant) and $\sigma$ is
one of the two spin projections; the factor of 2 comes from the
spin summation. The diagrammatic expansion of the numerator is similar
to that for $Z$, with the diagram weight given by
\begin{equation}
\mathcal{D}^{(\nu)}(\mathcal{S}_p)\;  =\; 2\,  (-U)^p \, \det
\mathbf{B}^{\sigma}_{p+1}(\mathcal{S}_p, \mathbf{x},\tau) \,
\det \mathbf{A}^{-\sigma}_{p}(\mathcal{S}_p)\; ,
\label{w-dens}
\end{equation}
Here $\mathcal{S}_p \in \mathfrak{S}^{(Z)}$,
$\mathbf{A}^{-\sigma}_p(\mathcal{S}_p)$ is a $p \times p$
matrix (\ref{matrix}), and
$\mathbf{B}^{\sigma}_{p+1}(\mathcal{S}_p, \mathbf{x},\tau)$ is a
similar $(p+1)\times (p+1)$ matrix with an extra row and a column,
corresponding to the extra creation and annihilation operators in
the numerator of (\ref{dens-gibbs}), respectively. This
immediately leads to the following estimator (\ref{esti-2}) for
the number density
\begin{equation}
\mathcal{Q^{(\nu)}}(\mathcal{S}_p)\; =\;  2\; \frac{ \det \mathbf{B}^
{\sigma}_{p+1}(\mathcal{S}_p, \mathbf{x},\tau) }%
 { \det \mathbf{A}^{\sigma}_p(\mathcal{S}_p) }\; \delta^{(Z)}(\mathcal
{S}_p)\;
 .
\label{R-dens}
\end{equation}
We utilize the freedom of choosing $(\sigma,\mathbf{x},\tau)$ to
suppress autocorrelations in measurements. The density measurement
starts with randomly generated $(\sigma,\mathbf{x},\tau)$.

The estimator for kinetic energy is derived similarly. One
employs the coordinate-space expression for the kinetic energy in
terms of hopping operators and deals with a slightly generalized
version of Eq.~(\ref{dens-gibbs}):
\begin{equation}
 \langle\,
c^{\dagger}_{\mathbf{x+g}, \sigma}\,  c_{\mathbf{x} \sigma}\,
\rangle \; =\;   \, \frac{ \Tr c^{\dagger}_{\mathbf{x+g},
\sigma}(\tau) \, c_{\mathbf{x} \sigma}(\tau) \, e^{-\beta H}
}{Z}\; .
\label{eps}
\end{equation}
The rest is identical to the previous discussion up to replacement
$\mathbf{B}^{\sigma}_{p+1}(\mathcal{S}_p, \mathbf{x},\tau) \; \to
\; \mathbf{B}^{\sigma}_{p+1}(\mathcal{S}_p, \mathbf{x},{\bf g},
\tau)$, since now the spatial position of the creation operator is
shifted from that of the annihilation operator by the
vector ${\bf g}$. In our case, only the nearest-neighbor
correlator (\ref{eps}) has to be computed.

\subsubsection{Estimator for the interaction energy}
\label{sssec:PE}

The estimator for the interaction energy
\begin{equation}
\langle H_1 \rangle \; =\; { \Tr  H_1 e^{-\beta H} \over Z}
\label{PE}
\end{equation}
is readily constructed using a generic trick of finding the
expectation value of operator in terms of which the
perturbative expansion is performed. Consider the Hamiltonian
$H(\lambda) = H_0 + \lambda H_1$ and observe that
\begin{equation}
\Tr  H_1 e^{-\beta H}\; =\; -{1\over \beta}\, {\partial \over
\partial \lambda} \, \Tr e^{-\beta H}
\; \equiv\;  -{1\over \beta}\, {\partial Z
\over
\partial \lambda}  \; .
\label{PE-identity}
\end{equation}
The differentiation of Eq.~(\ref{Z-diagr}) for $Z=Z(\lambda)$
and letting $\lambda =1 $ afterwards is straightforward since
the diagram of order $p$ is proportional to $\lambda^p$. Hence
\begin{equation}
\mathcal{Q}^{(H_1)}\left( \mathcal{S}_p\right)\;  =\; -
\beta^{-1}p\; \delta^{(Z)}(\mathcal{S}_p) \; .
\label{R-PE}
\end{equation}

\subsubsection{Estimator for the integrated correlation function}
\label{sssec:g-im}
Following the general treatment of Ref.\ \cite{polaron00}, one can
construct an estimator for the correlation function (\ref{corr}).
In this work we just need
to sum and integrate this correlator over all its variables (see
Sec.\ \ref{sec:sense}):
\begin{equation}
K(L,T)\;  =\;  (\beta L^d)^{-2} \sum_{\mathbf{x}, \mathbf{x}'}
\int_0^{\beta} \rmd\tau \int_0^{\beta}\rmd\tau'
G_2(\mathbf{x}-\mathbf{x}',\, \tau - \tau')\; ,
\label{Q-rescaled}
\end{equation}
and the estimator for $K(L,T)$ is particularly
simple:
\begin{equation}
\mathcal{Q}^{(K)}(\mathcal{S}_p)\;  =\;  (\beta L^d)^{-2}
\zeta^{-1} \delta^{(G)}(\mathcal{S}_p) \; . %
\label{R-g-im}
\end{equation}


\section{Extrapolation towards macroscopic continuum system}
\label{sec:sense}

The MC setup discussed in Sec.\ \ref{sec:ddmc} works in the grand
canonical ensemble with external parameters $\{L,T,\mu
\}$. In order to extract the critical temperature of a continuum
gas from MC data, one has to perform the two-step
extrapolation. (i) Upon taking the limit of $L\to \infty$ one
obtains $T_c(\mu)$, the critical temperature of a lattice system
at a given chemical potential, and translates it into $T_c(\nu)$ by
extrapolating the measured filling factor to the infinite system
size:
 $\nu\, \equiv\,  \nu(\mu,\, T=T_c(\mu), \, L\to\infty)$. (ii) The
extrapolation towards the continuum limit is then done by taking
the limit of $\nu\to 0$. The latter procedure is based on results
presented in Sec.\ \ref{sec:2body}.


The finite-size extrapolation is performed by considering
a series of system sizes $L_1 < L_2 < L_3 \,\dots$\, . At
the critical point the correlation function (\ref{corr}) decays at
large distances as a power-law:
$G_2(\mathbf{x}-\mathbf{x}',\, \tau - \tau') \propto
|\mathbf{x}-\mathbf{x}'|^{-(1+\eta)}$, where $\eta$ is the
anomalous dimension \cite{Fisher}. Since we expect the transition
to belong to U(1) universality class, we take $\eta \approx 0.038$. If one
rescales the summed correlator (\ref{Q-rescaled}) according to
\begin{equation}
R(L,T) = L^{1+\eta} K(L,T)\; , \label{K-rescaled}
\end{equation}
the corresponding quantity is supposed to become size-independent
at the critical point, i.e. the crossing of $R(L_i,T)$ and $R(L_j,T)$
curves can be used to obtain an estimate $T_{L_i,L_j}$
for the critical temperature $T_c(\mu)$ \cite{Binder81}.
Indeed, for temperatures in the
vicinity of the critical point the correlation length diverges as
$\xi_\mathrm{corr} \propto |t|^{-\nu_\xi}$, where
$t=(T_c(\mu)-T)/T_c(\mu)$, and $\nu_\xi\approx 0.671$ for the U(1)
universality class.
In the renormalization group (RG) framework \cite{Fisher}, the
finite-size scaling of the rescaled correlator $R$ obeys the
relation
\begin{equation}
R = f\left(x\right)(1 + cL^{-\omega}+\dots)\; ,
 \label{RG1}
\end{equation}
where $x=(L/\xi_\mathrm{corr})^{1/\nu_\xi}$ is the dimensionless
scaling variable, $f(x)$ is the universal scaling function
analytic at $x = 0$, $c$ is a non-universal constant,
$\omega\approx 0.8$ is the critical exponent of the leading
irrelevant field \cite{Zinn-Justin}, and dots represent
higher-order corrections to scaling. If the irrelevant field
corrections were not present, all $R(L_i,T)$ curves would
intersect at a unique point, $T_c(\mu)$.
Expanding Eq.\ (\ref{RG1}) up to terms linear in $t$ one
obtains for the crossing $T_{L_i,L_j}$
\begin{equation}
T_{L_i,L_j} - T_c(\mu)\;  =\;
\frac{\mathrm{const}}{L_j^{1/\nu_\xi+\omega}}%
\, \frac{\left( L_j/L_i \right)^{\omega} - 1 }%
{ 1 - \left( L_i/L_j \right)^{1/\nu_\xi} }\; .
\label{Tc-fit}
\end{equation}
To employ Eq.\ (\ref{Tc-fit}) one performs a linear fit of the
sequence of $T_{L_i,L_j}$ against the right hand side of Eq.\ (\ref{Tc-fit})
for several pairs of system sizes.
The intercept of the best-fit line yields the thermodynamic limit
critical temperature $T_c(\mu)$. Note that if the universality
class is not U(1) and the values of $\eta$, $\nu_{\xi}$, $\omega$
are different, or system sizes are too small to justify the scaling limit,
the whole procedure fails. Hence, the adopted scheme of pinpointing
$T_c$ features a built-in consistency check.

\section{Thermodynamic scaling functions of a unitary Fermi gas}
\label{sec:thermodynamics}


As has been noted in Sec.\ \ref{sec:intro}, the only relevant
microscopic energy scale in the continuum unitary fermi gas is
given by the Fermi energy $\varepsilon_F = \kappa\,
{\hbar^2 n^{2/3} / m}$, where $\kappa = (3\pi^2)^{2/3}/2$ for a
two-component Fermi gas. Therefore, as it was first noticed in
Ref.\ \cite{Ho04}, all thermodynamic potentials feature
self-similarity properties and can be written in terms of
dimensionless scaling functions of the dimensionless temperature
$x=T/\varepsilon_F $. All scaling functions are mutually
related; it is sufficient to know just one of them to
unambiguously restore the rest. Apart from the shape of scaling
functions, the self-similarity at the unitary point
is {\it identical} to that of a non-interacting gas
Fermi gas \cite{Vtom}, including functional relations
between different thermodynamic potentials.
A derivation of scaling functions and relations
between them can be found in Ref.\ \cite{Ho04}. In this section,
we render the scaling analysis in a form convenient for our MC
study.

In terms of the dimensionless chemical potential $y = \mu
/\varepsilon_F$, the dimensionless equation of state reads $y =
f_{\mu}(x)$. The $f_{\mu}$ function can be calculated numerically.
Another quantity which is also available in
our simulation is the dimensionless energy per particle ${E/(N
\varepsilon_F)}= f_E(x)$. The scaling relations for other
thermodynamic quantities are defined likewise. For instance,
the entropy and pressure read $ S/N = f_S(x)$, and $ {P/(n
\varepsilon_F)}=f_P(x) $. Though $f_S$ and $f_P$ are not
directly calculated in our simulation, and we can relate them
to $f_E$. It is also important to relate $f_{\mu}$ to
$f_E$, since this yields a consistency check for the
numerical results.

To establish desired relations, we start with the scaling relation
for the Helmholtz free energy ${F/(N \varepsilon_F)} = f_F(x)$
which in canonical variables reads
\begin{equation}
F(T,N,V) \; = \; \gamma \, f_F\left( T/\gamma(N/V)^{2/3}\right) \,
(N/V)^{2/3}\,  N \; , \label{}
\end{equation}
where  $\gamma = {\kappa \hbar^2  / m}$.  Then, for the entropy
and pressure we have (the prime stands for the derivative)
\begin{eqnarray}
f_S \; = \;
-f_F'\, , \label{S2} \\
f_P \; = \;
(2/3) (f_F - f_F'\, x) \; . \label{P2}
\end{eqnarray}
The expression for energy in terms of $f_F$ is
\begin{equation}
f_E\; =\; f_F - f_F'\, x  \; .
\label{E2}
\end{equation}
We thus see that
\begin{equation}
f_P\; \equiv \; (2/3)f_E \; . \label{P3}
\end{equation}
On may also consider Eq.~(\ref{E2}) as a differential equation:
\begin{equation} f_F - f_F'\, x \; = \; f_E \;  \label{rel0}
\end{equation}
to be integrated with respect to $f_F$ from $x = \infty$ down to
finite $x$, taking advantage of known asymptotic behaviour of
$f_F$ and $f_E$ for the weakly interacting two-component gas.

Now, we note that from the general thermodynamic relation
$E = -PV+TS+\mu N$
it immediately follows that
\begin{equation} f_E \; = \; -f_P+xf_S+f_{\mu} \; , \label{rel3}
\end{equation}
which, in turn, leads to the following relations
\begin{eqnarray}
f_S\; =\; {(5/3)f_E-f_{\mu} \over x} \; , \label{S3} \\
f_F\; =\; f_{\mu} -(2/3)f_E \; , \label{F2}
\end{eqnarray}
that allow one to express $f_S$ and $f_F$ functions through
numerically available functions  $f_E$ and $f_{\mu}$. Another
useful relation is
\begin{equation} f_F'\; =\; {f_{\mu} -(5/3)f_E\over x} \; , \label{F_prime}
\end{equation}
which allows to extract $f_F'$ directly from $f_E$ and $f_{\mu}$,
and thus provides a simple check for the data consistency:
The result (\ref{F2}) for the $f_F$ curve should be consistent with the
the derivative deduced from (\ref{F_prime}).

By integrating equation (\ref{rel0}) we get
\begin{equation}
f_F(x)\; =\; C_0\, x\, -\, {3\over 2}\, x\ln x\,  -\,
x\int_x^\infty \left ( {3\over 2}{1\over x_0}\,  -\,  {f_E \over
x_0^2}\right) \rmd x_0 \; . \label{F3}
\end{equation}
Here we took into account the asymptotic ideal-gas behaviour of
$f_E$:
\begin{equation}
f_E(x) \; \to \; {3\over 2}\, x~~~~~~~{\rm at}~~~~~~~x\; \to \;
\infty \; , \label{ass_E}
\end{equation}
and introduced the corresponding term into the integral to render
the latter convergent. The free constant of integration, $C_0$,
can be restored from (\ref{F2})-(\ref{ass_E}) combined with the
asymptotic ideal-gas behaviour of $f_\mu$,
\begin{equation}
f_\mu (x) \; \to \; -{3\over 2}\, x \ln \left( {\kappa \over 2
\pi}\, x \right )~~~~~~~{\rm at}~~~~~~~x\; \to \; \infty \; .
\label{ass_mu}
\end{equation}
The result is
\begin{equation}
C_0\; =\; {3\over 2}\ln \left( {2 \pi\over \kappa }  \right ) \,
-\, 1 \; . \label{C_0}
\end{equation}
Note that if higher-order terms in the asymptotic ($x\to \infty$)
behaviour of $f_E(x)$ are also known, then (\ref{F3}) can be used
to establish the corresponding corrections for $f_F(x)$ and other
scaling functions. For instance, it has been found in Ref.\
\cite{HoMueller04} that, as $x\to\infty$
\begin{equation}
f_E(x) \to {3\over 2}\, x  - {9\over 8}\,  \left(  {\pi \over
\kappa } \right)^{3/2}{1\over \sqrt{x}}  . \label{H_M_E}
\end{equation}
In accordance with (\ref{F3}), this implies ($x\to\infty$)
\begin{eqnarray}
f_F(x) \to C_0\, x  -  {3\over 2}\, x\ln x  - {3\over 4}\, \left(
{\pi \over \kappa } \right)^{3/2}{1\over \sqrt{x}}, \label{H_M_F}%
\\
f_\mu(x)  \to  -{3\over 2}\, x \ln \left( {\kappa \over 2 \pi}\, x
\right ) -   {3\over 2}\,  \left(  {\pi \over \kappa }
\right)^{3/2}{1\over \sqrt{x}}  . \label{H_M_mu}
\end{eqnarray}

Finally, the scaling functions $\mathcal{W}_0$ and
$\mathcal{G}_0$ defined in Ref.\ \cite{Ho04} are related to
$f_E$ and $f_{\mu}$ as follows:
\begin{eqnarray}
\mathcal{W}_0 \left(f_{\mu}(x)/x\right) \; \equiv\;
\frac{40}{9\sqrt{\pi}}\,
\frac{f_E(x)}{x^{5/2}}, \\
\mathcal{G}_0 \left(x/f_{\mu}(x)\right)  \; \equiv\;
\frac{5}{3}\, \frac{f_E(x)}{f_{\mu}(x)^{5/2}}\; .
\label{Ho-functions}
\end{eqnarray}

\subsection{Trapped Fermi gas \label{ssec:LDA}}

So far we have considered the uniform Fermi gas, while in
experimental realizations
\cite{BEC-BCS-expt,Hulet-PRL2005,Grimm-in-situ,Hulet-Science2006,expt-univ-energy,expt-ramp}
one has to deal with the parabolic trapping potential. The
standard procedure, especially in systems with short ``healing
length'' is to use the local density approximation (LDA),
\textit{i.e.} to replace the chemical potential with its
coordinate-dependent counterpart $\mu(\mathbf{r})=\mu -
V(\mathbf{r})$. This procedure can be easily combined with MC
results as follows. We introduce the dimensionless variable
$u=\mu/T=f_\mu(x)/x$ and define the scaling function for the
number density as $w_{n}=x^{-3/2}$. This is equivalent to the
parametric $\{ u(x), w_n(x)\}$ dependence of $n$ on $u$. The
scaling functions for other thermodynamic quantities are defined
in a similar manner, e.g. $w_{E}(u) \equiv f_E(x(u))$ for the
energy, and $w_S \equiv f_S (x(u))$ for the entropy. Within LDA
$u$ acquires the coordinate dependence
$u(\mathbf{r})=(\mu-V(\mathbf{r}))/T$ which translates into the
density profile $n(\mathbf{r};\mu,T) = w_n(u(\mathbf{r}))
(mT/\kappa\hbar^2)^{3/2}$. Likewise, other thermodynamic functions
are to be understood as local, coordinate-dependent quantities.

Consider the case of $N$ particles in  a cigar-shaped parabolic
trap, characterized by the axial and radial frequencies
$\omega_\|$ and $\omega_\perp$. The characteristic energy in this
case is $E_F = (3N)^{1/3} \hbar (\omega_\|^2\omega_\perp)^{1/3}$,
which would coincide with the Fermi energy for a non-interacting
gas in the trap. Note, that we denote the Fermi energy in the trap
by an upright capital $E_F$ in order to avoid confusion with the
uniform system Fermi energy $\varepsilon_F$.

By integrating over the radial coordinates one obtains the axial
density profile $n_a(z)$ ($z$ is an axial coordinate) in the form
\begin{equation}
\frac{n_a(z)}{N} = \frac{(2T/E_F)^{5/2}}{\pi L_\|}
\overline{w}_n\left( \frac{\mu}{T} - \frac{z^2}{2(T/E_F)L_\|^2}
\right),
\label{n_a}
\end{equation}
where $L_\| = \lambda^{1/3} (3N)^{1/6} l_\|$, the aspect ratio
$\lambda=\omega_\perp / \omega_\|$, the oscillator length $l_\|^2=
\hbar/m\omega_\|$, and
\begin{equation}
\overline{w}_n(p) = \int_{-\infty}^{p} w_n(u) \rmd u.
\label{wbar}
\end{equation}

By integrating Eq.\ (\ref{n_a}) with respect to $z$, one finally
relates
 chemical potential to temperature:
\begin{equation}
\overline{\overline{w}}_n\left(\frac{\mu}{T}\right) \left(
\frac{T}{E_F} \right)^3 = \frac{\pi}{16},
\label{wbarbar}
\end{equation}
where
\begin{equation}
\overline{\overline{w}}_n\left( p \right) = \int_{0}^{\infty}
\overline{w}_n(p-q^2) \rmd q.
\label{wbarbar_def}
\end{equation}

Obtaining the temperature dependence of thermodynamic functions
for a non-uniform system within LDA is also straightforward. For the
total energy of a cloud $E_\mathrm{tot}$ we obtain
\begin{equation}
\frac{E_\mathrm{tot}}{N E_F} = \frac{16}{\pi}\left(
\frac{T}{E_F}\right)^4 \int_{-\infty}^{\mu/T} \rmd p
\int_{0}^{\infty} \rmd q\, w_E(p-q^2) w_n^{5/3}(p-q^2),
\label{Etot}
\end{equation}
and likewise for the entropy:
\begin{equation}
\frac{S}{N} = \frac{16}{\pi}\left( \frac{T}{E_F}\right)^3
\int_{-\infty}^{\mu/T} \rmd p \int_{0}^{\infty}\rmd q\, w_S(p-q^2)
w_n(p-q^2).
\label{Stot}
\end{equation}
%

\section{Results and Discussion}
\label{sec:results}

We performed simulations outlined in previous
Sections for filling factors ranging from $0.95$ down to $0.06$
with up to about 300 fermions on lattices with up to $16^3$
sites. The typical rank of determinants involved in computations
of acceptance ratios (Sec.\ \ref{sec:updates}) and estimators
(Sec.\ \ref{ssec:estimators}) is up to $M \sim 5000$. Since we
only need ratios of determinants, we use fast-update formulas
\cite{Rubtsov} to reduce the computational complexity of updates
from $M^3$ down to $M^2$.

We validate our numerical procedure by comparing results against
the exact diagonalization data for a $4\times 4$ cluster
\cite{Husslein}, and other simulations of the critical temperature at
quarter filling $\nu=0.5$ \cite{Zotos,DCA} and $\nu=0.25$
\cite{DCA}. In all cases we find agreement within statistical
errors of a few percent.

\subsection{Critical temperature}
\label{ssec:tc}

\begin{figure}
\includegraphics[width=0.75\columnwidth,keepaspectratio=true]{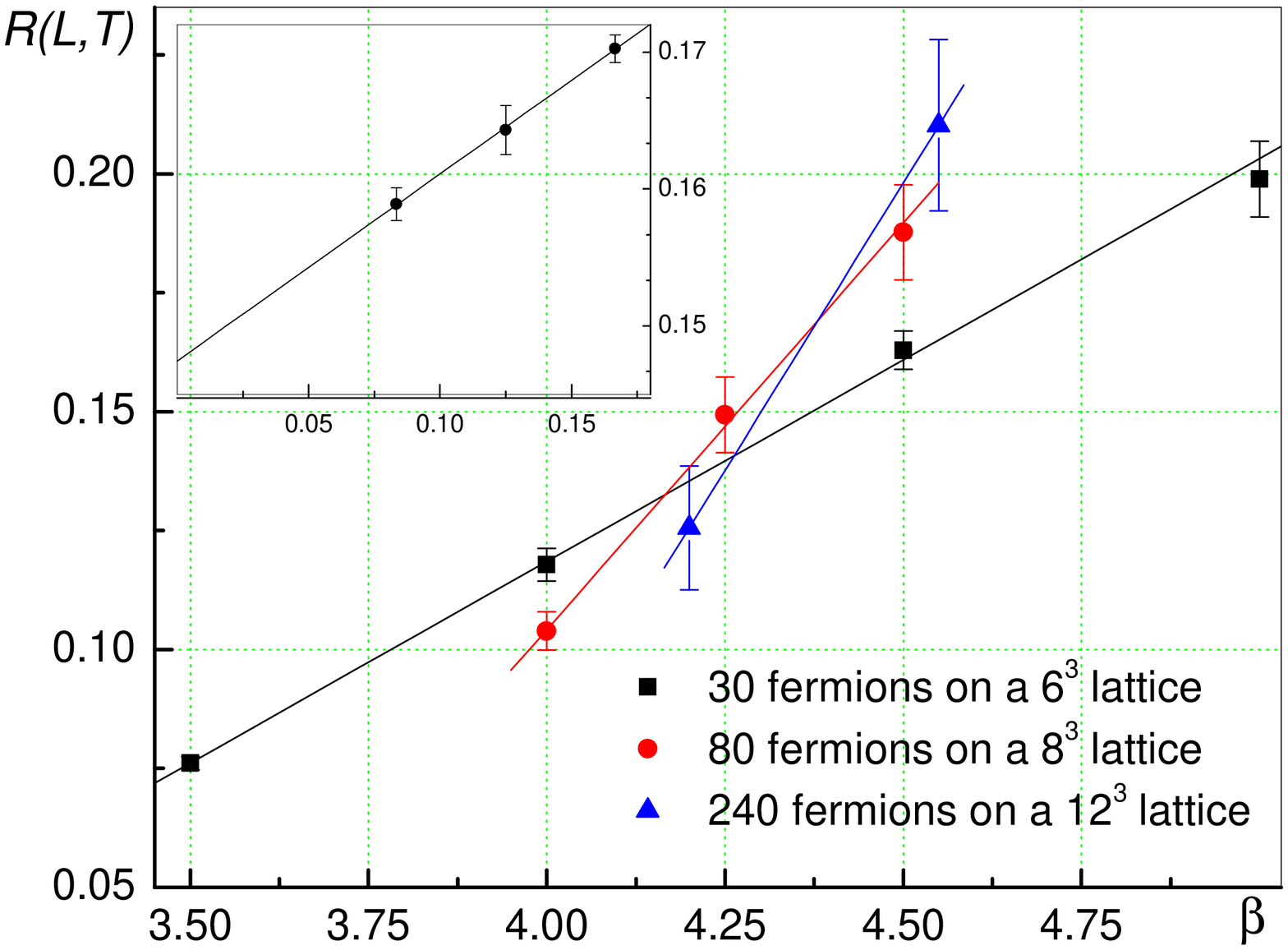}
\caption{ A typical crossing of the $R(L,T)$ curves.  The
errorbars are $2\sigma$, and solid lines are the linear fits to
the MC points. Inset shows the finite-size scaling of the filling
factor ($\nu$ vs $1/L$), which yields $\nu=0.148(1)$. From this
plot and Eq.\ (\ref{Tc-fit}) one obtains $1/T_c(\nu) = 4.41(5)/t$
}.
\label{fig:crossing}
\end{figure}

Figure \ref{fig:crossing} shows a typical example of the
finite-size analysis outlined in Sec.\ \ref{sec:sense}. Despite
the fact that numerically accessible system sizes are quite small,
\fref{fig:crossing} (and similar analysis for the whole range of
filling factors) supports expectations that the universality class
for the SF phase transition is U(1).  The finite-size analysis
allows us to pinpoint the phase transition temperature to within a
few percent.

\begin{figure}
\includegraphics[width=0.75\columnwidth,keepaspectratio=true]{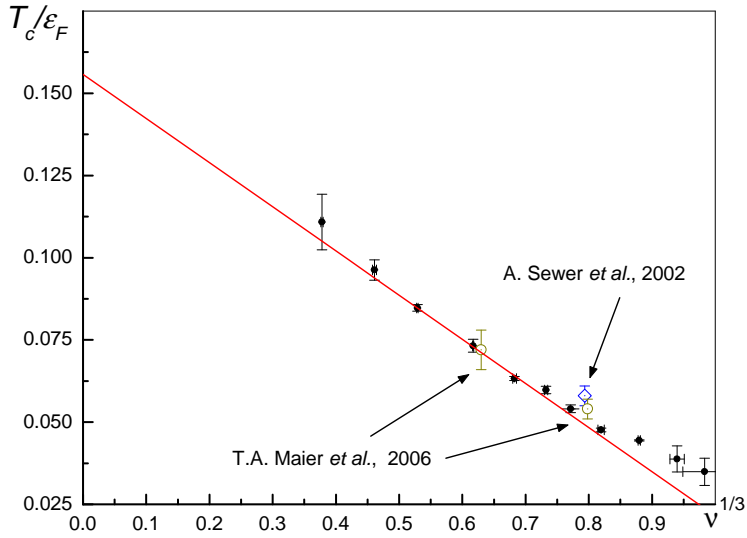}
\caption{ The scaling of the lattice critical temperature with
filling factor (circles). $\nu=1$ corresponds to the half filling.
The errorbars are one standard deviation. The results of Ref.\
\cite{Zotos,DCA} at quarter filling and $\nu=0.25$ are also shown
for a comparison. See the text for discussion.  }
\label{fig:tc}
\end{figure}

Shown in \fref{fig:tc} is the dependence of the critical
temperature on the lattice filling factor. The critical
temperature is measured in units of the Fermi energy, as is
natural for the unitarity limit. We define the Fermi momentum for
a lattice system with filling factor $\nu$ as $k_F = (3\pi^2
\nu)^{1/3}$ and the Fermi energy $\varepsilon_F = k_F^2$, as those
of a continuum gas with the same effective mass and number density
$n=\nu$.

It is clearly seen that the presence of the lattice suppresses the
critical temperature considerably, nearly by a factor of $4$,
depending on the filling factor. Strong dependence of $T_c$
on $\nu$ is in apparent contradiction with Ref. \cite{Bulgac},
which claims weak or no $\nu$-dependence. This disagreement
might be due to the difference in the single-particle spectra
$\varepsilon_\mathbf{k}$ used: Ref.\ \cite{Bulgac} employs the
parabolic spectrum with spherically symmetric cutoff, while we use
the tight-binding dispersion law. Indeed, Eqs.\
(\ref{Gamma-A})-(\ref{Gamma-B}) indicate that a particular choice
of $\varepsilon_\mathbf{k}$ does influence lattice corrections
to $T_c$, which may even have different signs for different
$\varepsilon_\mathbf{k}$.

\begin{figure}
\includegraphics[width=0.75\columnwidth,keepaspectratio=true]{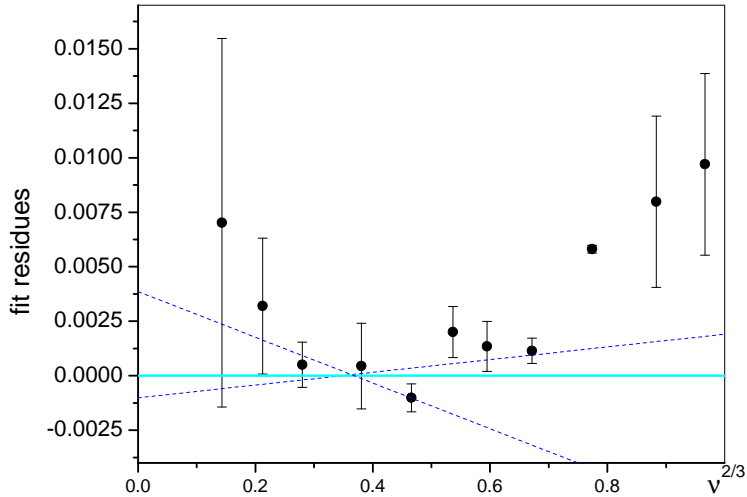}
\caption{ The fit residues for the best-fit line of \fref{fig:tc},
plotted versus $\nu^{2/3}$ (circles). Zero level is shown by the
horizontal line, the blue dashed lines are linear fits to the data
points for $\nu<0.5$ and $\nu<0.35$, respectively. }
\label{fig:residues}
\end{figure}

It is also clear from \fref{fig:tc} that close to half-filling
$T_c$ is essentially constant, as expected (see, \textit{e.g.}
\cite{Micnas90}). The predicted $\sim \nu^{1/3}$ scaling
(\ref{nu-13}) sets in at about $\nu \approx 0.5$. We thus use a
linear fit $T_c(\nu)/\varepsilon_F(\nu) = T_c/\varepsilon_F -
\mathrm{const}\cdot \nu^{1/3}$ to eliminate lattice corrections in
the final result. Such fitting procedure results in the best-fit
line given by $0.152(7) - 0.13(2)\nu^{1/3}$. We further analyze
the fit residues in order to estimate the effect of the
sub-leading lattice corrections which are expected to be
proportional to $\nu^{2/3}$. As shown in \fref{fig:residues}, such
corrections, if any, are smaller than the uncertainty of the
$\nu^{1/3}$ fit.

This analysis yields the final result $T_c/\varepsilon_F =
0.152(7)$ for the continuum uniform gas, which is noticeably below
the transition temperature in the BEC limit $T_\mathrm{BEC} =
0.218 \varepsilon_F$. Various approximate analytical treatments
led in the past to $T_c$ either above
\cite{NSR,Randeria95,Holland-Timmermans,Perali}, or below
\cite{Haussmann94,Ohashi-Griffin,LiuHu} $T_\mathrm{BEC}$.

Is is instructive to compare our results for $T_c$ to other
numerical calculations available from the literature. The
simulations of Ref. \cite{Wingate} yield $T_c = 0.05
\varepsilon_F$, but at the value of the scattering length which
has not been determined precisely. This result most probably
corresponds to a deep BCS regime, where the transition temperature
is exponentially suppressed. Lee and Sch\"{a}fer
\cite{Lee-Schaefer} report an upper limit $T_c < 0.14
\varepsilon_F$, based on a study of the caloric curve of a unitary
Fermi gas down to $T/\varepsilon_F=0.14$ for filling factors down
to $\nu=0.5$. The caloric curve of Ref.\ \cite{Lee-Schaefer} shows
no signs of divergent heat capacity which would signal the phase
transition. This upper limit is consistent with
$T_c(\nu=0.5)/\varepsilon_F \approx 0.054$, see \fref{fig:tc}.

\begin{figure}
\includegraphics[width=0.75\columnwidth,keepaspectratio=true]{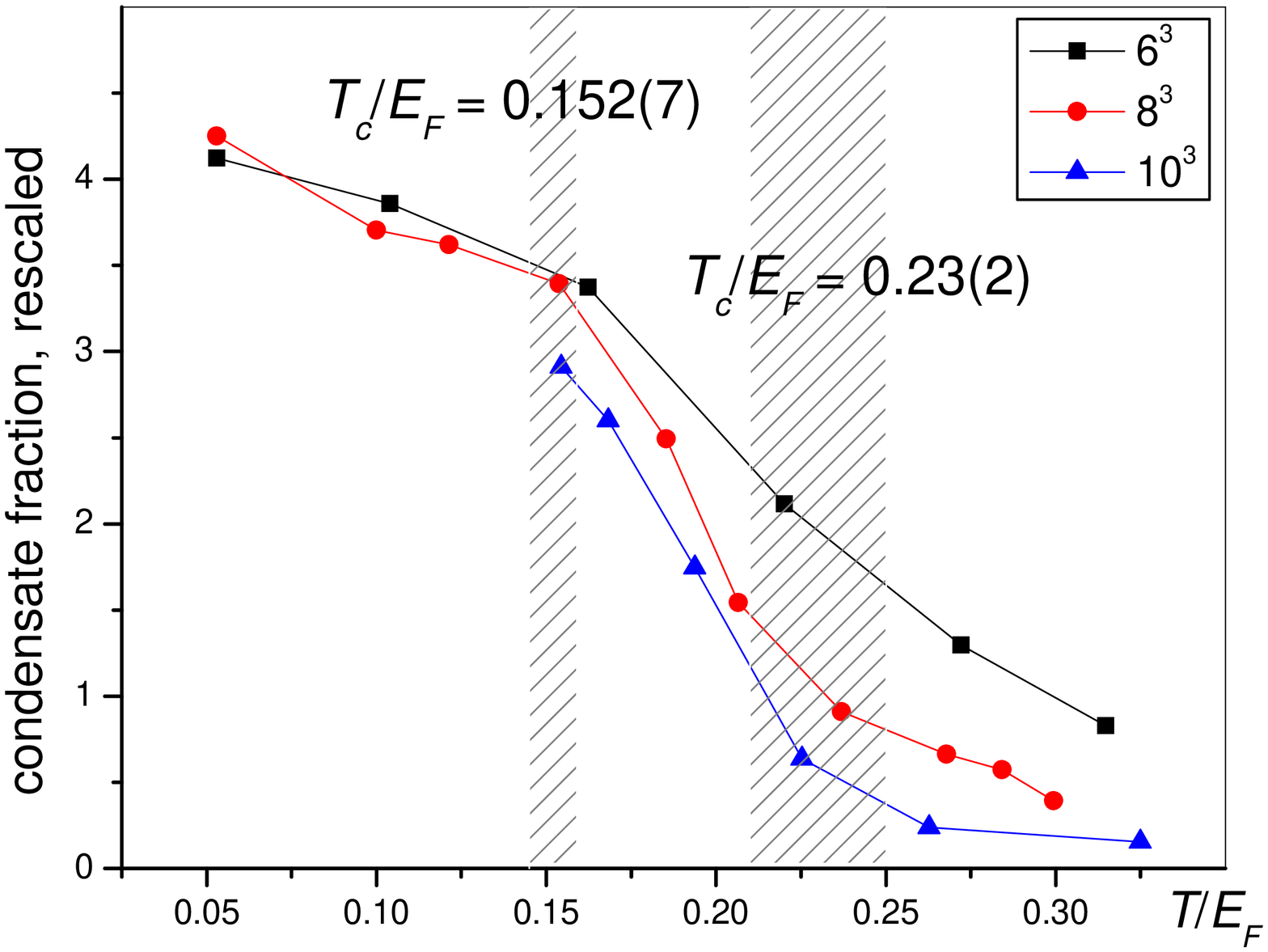}
\caption{ The finite-size scaling of the condensate fraction data
from Ref.\ \cite{Bulgac}. Raw data points are rescaled similar to
Eq.\ (\ref{K-rescaled}) by the $L^{1+\eta}$ factor. Shaded
vertical strips represent results for $T_c/\varepsilon_F$ of this
work and Ref.\ \cite{Bulgac}, respectively, solid lines are drawn
to guide an eye.}
\label{fig:crossing-b}
\end{figure}

The Seattle group has performed simulations of the caloric curve
and condensate fraction, $n_0$, of the unitary gas, Ref.\
\cite{Bulgac}. Using ``visual inspection'' of the caloric curve
shape the critical temperature was estimated in Ref.\
\cite{Bulgac} to be $T_c=0.22(3)\varepsilon_F$. Unfortunately, the
authors did not perform the finite-size analysis and $\nu \to 0$
extrapolation. The overall shape of the caloric curve seem to be
little affected by the finite volume of the system. This is hardly
surprising since even in the thermodynamic limit $E(T)$ and its
derivative $\rmd E/\rmd T$ are continuous at the transition point.
These properties also make it hard to use non-quantitative
measures for reliable estimates of critical parameters from the
$E(T)$ curve. On the other hand, the condensate fraction which has
singular properties at $T_c$ does show sizable finite-size
corrections, see figure~1 of Ref.\ \cite{Bulgac}. At this point we
note that scaling of the condensate fraction is identical to that
for $K(L,T)$. In \fref{fig:crossing-b} we plot the data of
Seattle's group as $n_0 L^{1+\eta}$ versus temperature. The
intersection of scaled curves turns out to be  inconsistent with
the estimate for $T_c$ derived from the caloric curve inspection.

\subsection{Thermodynamic functions}
\label{ssec:res:therm}

\begin{figure}
\includegraphics[width=0.99\columnwidth,keepaspectratio=true]{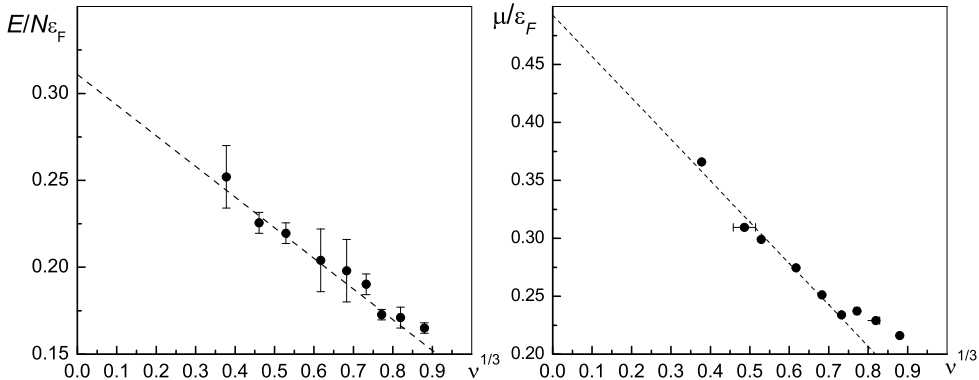}
\caption{ Energy (left-hand panel) and chemical potential
(right-hand panel) dependence on the filling factor along the
critical isotherm $T=T_c(\nu)$. Dots are the MC results, dashed
lines are the linear fits. }
\label{fig:emu-tc}
\end{figure}

The filling factor dependence of thermodynamic quantities is
similar to that of $T_c$: Figure~\ref{fig:emu-tc} displays the
behaviour of energy and chemical potential along the critical line
$T=T_c(\nu)$.  The extrapolation towards $\nu\to 0$ yields for the
continuum gas
\begin{eqnarray}
&E / (N\varepsilon_F) \; =\;  0.31(1)~~~~~~~~~ &(T=T_c)\; , \\
&\mu / \varepsilon_F\;   =\;  0.493(14)~~~~~~~~~&(T=T_c) \; .
\label{magic_numbers}
\end{eqnarray}
The numerical values for other thermodynamic functions at
criticality can be easily restored using the formulas of Sec.\
\ref{sec:thermodynamics}.

\begin{figure}
\includegraphics[width=0.75\columnwidth,keepaspectratio=true]{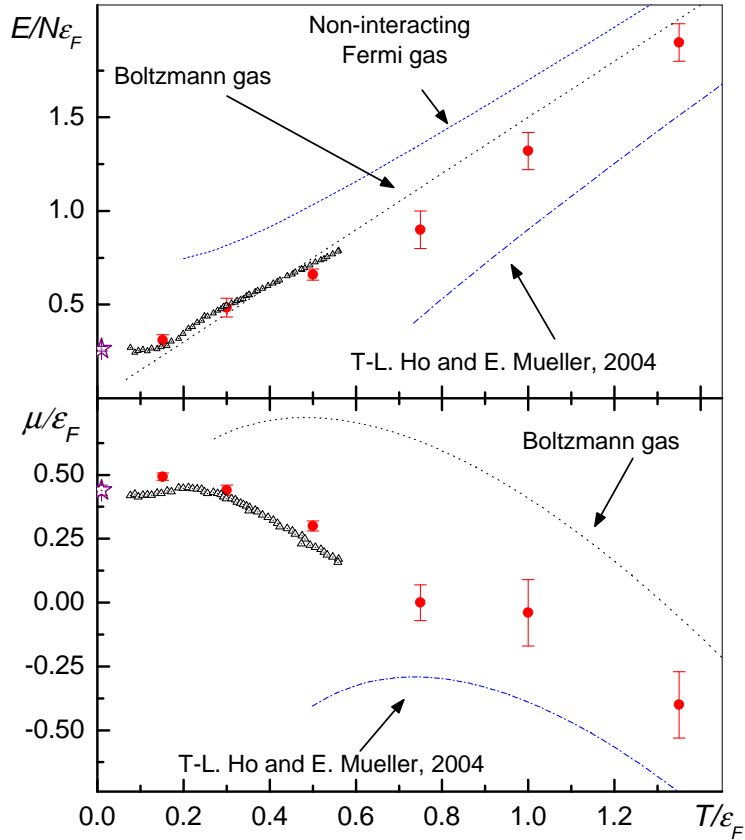}
\caption{ The temperature dependence of the energy per particle
(upper panel) and chemical potential (lower panel) of the unitary
Fermi gas. Red circles are the MC results, black dotted lines and
blue dashed lines correspond to the Boltzmann and non-interacting
Fermi gases, respectively, the dot-fashed lines are the asymptotic
prediction of Ref.\ \cite{HoMueller04} (plus the first virial
Fermi correction), black triangles are the MC results of Ref.\
\cite{Bulgac}, and the purple stars denote the ground-state
fixed-node MC results \cite{Giorgini-Carlson}. }
\label{fig:emu}
\end{figure}

In order to elucidate the thermodynamic behaviour of the unitary
gas, we performed simulations for a range of temperatures $T>T_c$.
Shown in \fref{fig:emu} are the simulation results for energy and
chemical potential for the continuum gas as functions of
temperature. Each point was obtained using data analysis similar
to that depicted in \fref{fig:emu-tc}. In the high-temperature
region we simulated up to $80$ fermions on lattices with up to
$32^3$ sites. In this region, the condition $\nu \ll 1$ is
necessary but not sufficient for extrapolation to the continuum
limit, for it is crucial to keep temperature much smaller than the
bandwidth: $T \ll 6t$.

As can be seen from \fref{fig:emu}, our results for both energy
and chemical potential approach the virial expansion
\cite{HoMueller04} as $T/\varepsilon_F \to \infty$. For
$T/\varepsilon_F \leqslant 0.5$ our data are not far from the
curve of Ref.\ \cite{Bulgac}. Though we do not have data points
for $T<T_c$ there is still a reasonable agreement even at $T_c$
with the $T\to 0$ fixed-node MC values \cite{Giorgini-Carlson}. In
this region, our results are consistent with a very weak
dependence of energy and chemical potential on temperature, and
the numeric values of both are consistent with the experimental
results \cite{Grimm-in-situ,Hulet-Science2006,expt-univ-energy}.


\begin{figure}
\includegraphics[width=0.75\columnwidth,keepaspectratio=true]{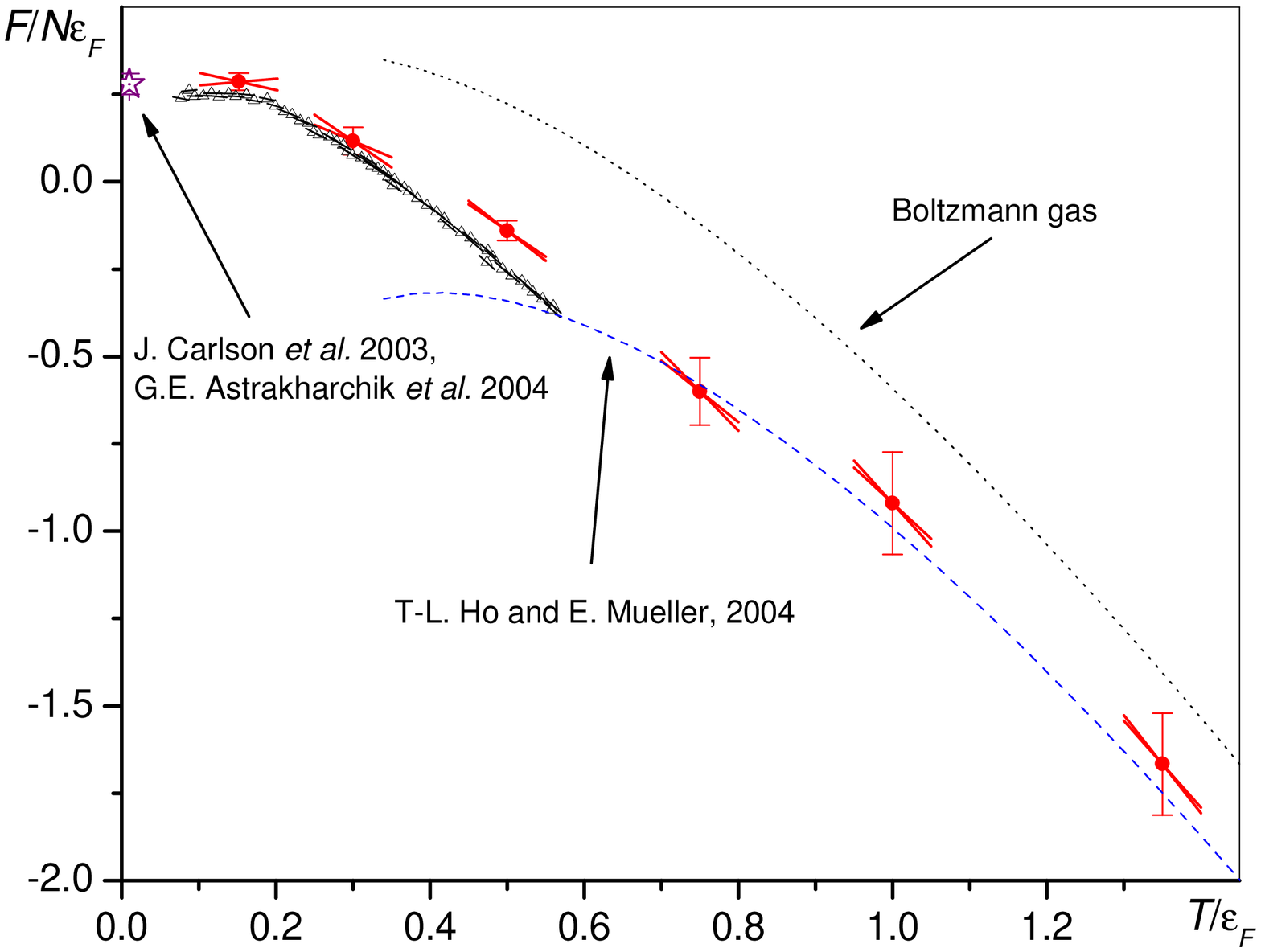}
\caption{ Free energy versus temperature. Red dots are the MC
data, and dashes represent the errorbar range for derivative of
free energy, calculated via Eq.\ (\ref{F_prime}). Black triangles
are MC results of Ref.\ \cite{Bulgac}, purple star denotes the
ground-state fixed-node MC result \cite{Giorgini-Carlson}, black
dotted line shows the Boltzmann gas curve, and the blue dashed is
the asymptotic prediction of Ref.\ \cite{HoMueller04}.}
\label{fig:consi}
\end{figure}

Using Eq.\ (\ref{F2}) and data from \fref{fig:emu} we deduce the
dependence of free energy on temperature, see \fref{fig:consi}. We
also use  Eq.\ (\ref{F_prime}) to make sure that our MC data for
energy and chemical potential are consistent with each other.

\subsection{Trapped gas\label{ssec:therm}}

As discussed in Sec.\ \ref{ssec:LDA}, the thermodynamic functions
for the uniform case can be used for analysis of experimental
system within the local density approximation.
In this section we report preliminary results of
our ongoing study of trapped gas experiments.

It starts with the interpolation procedure which produces
continuous functional behaviour for thermodynamic functions,
consistent with the discrete set of simulated points. We use a
piecewise-cubic ansatz with a smooth crossover to the virial
expansion, Eq.\ (\ref{H_M_F}), for the free energy. Temperature
dependence of both energy and entropy are then deduced using
numerical integration of Eqs.~(\ref{Etot}) and (\ref{Stot}),
respectively.

As the trapped gas is cooled down, the superfluidity first sets in
at the centre of the trap, where the density is the highest.
Equation (\ref{wbarbar}) can be used to pinpoint this onset
temperature: at $T=T_c$, $\mu/T=(\mu/\varepsilon_F^{0}) /
(T/\varepsilon_F^{0})$, where $\varepsilon_F^{0}$ is the Fermi
energy of the uniform gas with the density equal to the density at
the trap centre. Using $T_c/\varepsilon_F^{0} = 0.152(7) $ and Eq.
(\ref{magic_numbers}), one obtains $T_c/E_F = 0.20(2)$. We quote
here a conservative estimate for the uncertainty, which
incorporates both the uncertainty of the critical temperature
itself, and a systematic uncertainty which stems from restoring
the continuous functional dependence of the chemical potential out
of the finite set of the Monte Carlo calculated points with finite
errorbars.

Experimentally, the temperature of the strongly interacting Fermi
gas is not easily accessible. On the contrary, thermometry of the
non-interacting Fermi gases is well established. In the adiabatic
ramp experiments one starts from the non-interacting gas at some
temperature [in units of Fermi energy] $\left( T/T_F \right)
^{0}$, and slowly ramps magnetic field towards the Feshbach
resonance \cite{expt-ramp}, thus adiabatically connecting the
system at unitarity to a non-interacting one. Assuming the entropy
conservation during the magnetic field ramp, equation (\ref{Stot})
can be employed for the thermometry of the interacting gas: by
matching the entropy of a non-interacting gas at the temperature
$\left( T/T_F \right) ^{0}$ with the entropy calculated via Eq.\
(\ref{Stot}) one relates the initial temperature (before the
magnetic field ramp) to the final temperature (after the ramp). We
find that the onset of the superfluidity corresponds to $\left(
T/T_F \right) ^{0} = 0.12 \pm 0.02$ (again, we quote here the most
conservative estimate for the errorbar). This value seems to be
somewhat lower than the value suggested by the experimental
results \cite{expt-ramp}. Nevertheless, given the level of the
noise in figure 4 of Ref.\ \cite{expt-ramp} the consistency is
reasonable.

An alternative thermometry can be build on recent advances in the
experimental technique \cite{Grimm-in-situ,Hulet-Science2006}
which made it possible to directly image the \textit{in situ}
density profiles of the interacting system. Such density profiles
can be directly fit to Eq.\ (\ref{n_a}), which gives the shape of
the cloud depending on $\mu/T$ and $T/E_F$. By relating the
chemical potential to the temperature using Eq.\ (\ref{wbarbar}),
one is left with only one fitting parameter, $T/E_F$ (apart from a
trivial fitting parameter $z_0$ which accounts for the overall
shift of the cloud image off the trap centre).

\begin{figure}[th]
\includegraphics[width=0.75\columnwidth,keepaspectratio=true]{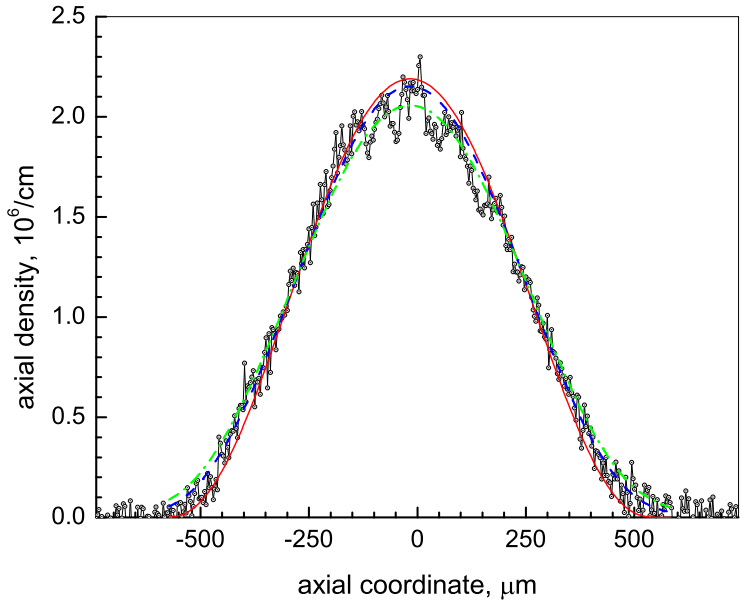}
\caption{ Axial density density profiles: experimental data (dots)
is from the data contained in figure 3 of
\cite{Hulet-Science2006}. The full red line is calculated via Eq.\
(\ref{wbar})--(\ref{wbarbar}) with $T/E_F = 0.03$
(correspondingly, $T/\varepsilon_F^{(0)}=0.02$), dashed blue line
corresponds to $T/E_F = 0.16$ ($T/\varepsilon_F^{(0)}=0.1$), and
dot-dashed green line is for $T/E_F = 0.22$
($T/\varepsilon_F^{(0)}=0.16$). In both cases we allowed for a
horizontal displacement of a whole curve in the range of $|z_0| <
20 \mu$m. See the text for discussion.}
\label{fig:fit}
\end{figure}

As an illustrative example of such procedure we have analyzed the
experimental density profiles measured by the Rice's group
\cite{Hulet-Science2006}, as depicted in \fref{fig:fit}. From this
analysis we deduce the upper bound for the temperature in the
experiments \cite{Hulet-Science2006} $T<0.1 E_F$, which is
consistent with the results of the measurements of the condensate
fraction \cite{Hulet-PRL2005}. Since in the experiments
\cite{Hulet-PRL2005,Hulet-Science2006} the gas is very degenerate,
one is able to put an upper limit on temperature only.

Note that if the temperature is known from, e.g.  the adiabatic
ramp experiments, Eqs.\ (\ref{n_a})-(\ref{wbarbar}) must reproduce
the cloud shape \textit{without free parameters} (apart from
$z_0$).

\section{Conclusions}
\label{sec:concl}

We have developed a worm-type scheme within the
systematic-error-free determinant diagrammatic Monte Carlo
approach for lattice fermions. We applied it to the Hubbard model
with attractive interaction and equal number of spin-up and
spin-down particles. At finite densities, the model describes
ultracold atoms in optical lattice. In the limit of vanishing
filling factor, $\nu \to 0$, and  fine-tuned (to the resonance in
the two-particle $s$-wave channel) on-site attraction, a universal
regime sets in, which is identical to the BCS-BEC crossover in the
continuous space. In the present work, we confined ourselves to a
special value of the on-site interaction, $U=U_*\approx -7.915 t$,
corresponding to the divergent $s$-scattering length. At $U=U_*$
and $\nu \to 0$, the system reproduces the unitary point of the
BCS-BEC crossover. The unitary regime is scale-invariant: all
thermodynamic potentials are expressed in terms of dimensionless
scaling functions of the dimensionless ratio $T/\varepsilon_F$
(temperature in units of Fermi energy). We obtained these scaling
functions by extrapolating results for the Hubbard model to $\nu
\to 0$. For the critical temperature of the superfluid-normal
transition in the uniform case we found
$T_c/\varepsilon_F=0.152(7)$. Our results form a basis for an
unbiased thermometry of trapped fermionic gases in the unitary
regime: In particular, we found (within the local density
approximation) that for the parabolic confinement, the critical
temperature in units of the characteristic trap energy $E_F$ is
$T/E_F = 0.20(2)$. For the experimentally relevant case of an
isentropic conversion of a  gas from the non-interacting regime to
the unitary regime we find that the onset of the superfluidity
corresponds to the initial temperature (before the magnetic field
ramp) $\left( T/T_F \right)^0 = 0.12\pm 0.02$, which is reasonably
consistent with the experimental result \cite{expt-ramp}, to
within the experimental noise.

\ack

 We appreciate generosity of A.~Bulgac, P.~Magierski, and
J.~Drut, who kindly provided us with their numeric data. We are
also indebted to W.~Li and R.~Hulet for providing us with their
unpublished experimental data. This research was enabled by
computational resources of the Center for Computational Sciences
and in part supported by the Laboratory Research and Development
program at Oak Ridge National Laboratory. Part of the simulations
were performed on the ``Hreidar'' cluster of ETH Z{\"u}rich. We
also acknowledge partial support by NSF grants Nos. PHY-0426881
and PHY-0456261.


\appendix

\section{Updating procedures} \label{sec:updates}

The generic diagrammatic MC rules for constructing updates are
as follows \cite{PST98}.  Let an update $\mathcal{B}$ transform a
diagram $\mathcal{D}(\mathcal{S}_p)$ into diagram
$\mathcal{D}(\mathcal{S}'_q)$. Here configurations
$\mathcal{S}_p$ and $\mathcal{S}'_q$ may have different order and
may or may not feature the ``worm'' two-point vertices.   The
update $\mathcal{B}$ involves two steps. First, a modification of
the configuration is proposed, with some probability density
$W(\mathcal{S}_p \to \mathcal{S}'_q)$ for new continuous/discrete
variables. There are no strict
requirements fixing the form of $W(\mathcal{S}_p \to
\mathcal{S}'_q)$, it should rather be chosen on physical grounds
to maximize the efficiency of the algorithm, and be simple
enough to allow analytical evaluation of the normalization integral.
Then, the update is either accepted, with probability $P_{\mathcal{S}_p \to
\mathcal{S}'_q}$, or rejected. The complimentary update
$\mathcal{C}$, which transforms $\mathcal{D}(\mathcal{S}'_q)$ into
$\mathcal{D}(\mathcal{S}_q)$ proposes the modification with the
probability density $W(\mathcal{S}_p \leftarrow \mathcal{S}'_q)$,
which, in principle, may differ from $W(\mathcal{S}_p \to
\mathcal{S}'_q)$, and is accepted with the probability
$P_{\mathcal{S}_p \leftarrow \mathcal{S}'_q}$. For a pair of
updates $\mathcal{B}$ and $\mathcal{C}$ to be balanced, the
acceptance probabilities must obey the Metropolis
relations \cite{PST98}
\begin{eqnarray}
\label{Metropolis_1}
P_{\mathcal{S}_p \to \mathcal{S}'_q} & = \min (1, \mathcal{R}),\\
P_{\mathcal{S}_p \leftarrow \mathcal{S}'_q} & = \min (1,
1/\mathcal{R})%
\label{Metropolis_2}
\end{eqnarray}
%
%
where $\mathcal{R}$ is a solution of the detailed-balance equation
\begin{equation}
\mathcal{R} W(\mathcal{S}_p \to \mathcal{S}'_q)
\mathcal{D}(\mathcal{S}_p) = W(\mathcal{S}_p \leftarrow
\mathcal{S}'_q) \mathcal{D}(\mathcal{S}'_q). %
\label{balance-eq}
\end{equation}

If the probabilities of selecting updates $\mathcal{B}$ and
$\mathcal{C}$ are not equal, they must also be included into
the definition of
$W(\mathcal{S}_p \to \mathcal{S}'_q)$ and $W(\mathcal{S}_p
\leftarrow \mathcal{S}'_q)$, respectively.

\subsection{The ``Diagonal'' updating scheme}

The minimal ergodic set of updates, as suggested in \cite{Rubtsov,
wePRB}, consists of just one pair of ``diagonal'' updates which
increase/decrease the number of vertices in a configuration by
one: in the forward update one adds a vertex at randomly selected
point in the $L^3 \times \beta$ hypercube; in the reverse update
one removes a random vertex from the configuration. In the
simplest version, the probability density for selecting a point is
uniform, i.e. one selects a particular lattice site with the
probability $p(\mathbf{x}_\mathrm{new})= L^{-d}$, and a temporal
position with the probability density $w(\tau_\mathrm{new}) =
\beta^{-1}$. A vertex to be removed is selected at random out of
$p$ vertices present in the configuration. Hence the Metropolis
acceptance ratio function (\ref{balance-eq}) is given by
\begin{equation}
\mathcal{R}_\mathrm{add} = \left| \frac{\det
\mathbf{A}(\mathcal{S}_{p+1})}{ \det \mathbf{A}(\mathcal{S}_{p}) }
\right|^2 \frac{(-U)\beta L^d}{p+1}\; .
\label{R-a/d}
\end{equation}
Here $\mathcal{S}_{p+1} =
(\mathbf{x}_\mathrm{new},\tau_\mathrm{new}) \cup \mathcal{S}_p $
is the configuration with an extra vertex
$(\mathbf{x}_\mathrm{new},\tau_\mathrm{new})$. The acceptance ratio
to remove a vertex is
\begin{equation}
\mathcal{R}_\mathrm{rem} = \left| \frac{\det
\mathbf{A}(\mathcal{S}_{p-1})}{ \det \mathbf{A}(\mathcal{S}_{p}) }
\right|^2 \frac{p}{(-U)\beta L^d}\; .
\label{R-a/d2}
\end{equation}

\subsection{Worm-type updating scheme}

\begin{figure*}
\includegraphics[width=0.99\textwidth,keepaspectratio=true]{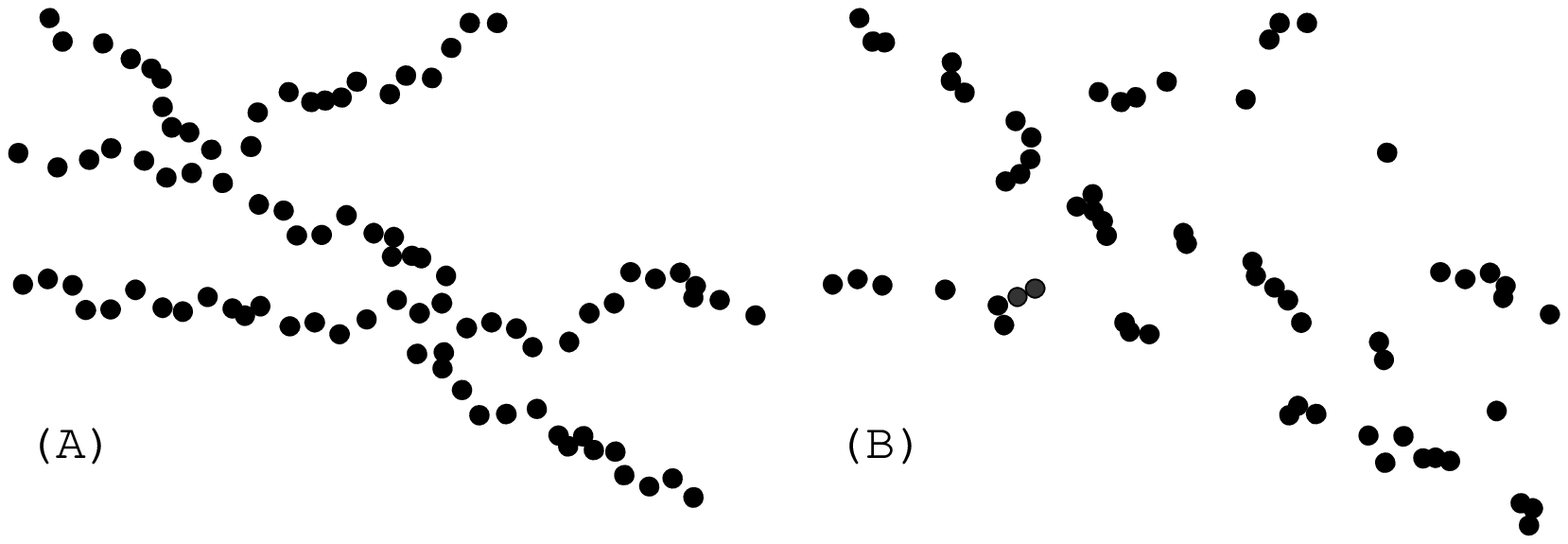}%
\caption{ Multi-ladder diagrams in deep BEC regime (A), and close
to unitarity (B). }
\label{fig:braids}
\end{figure*}

The diagonal updating scheme described in the previous section is
highly inefficient in the dilute regime---the regime we are mostly
interested in. To assess the efficiency of the ``diagonal'' scheme
consider first a low-density gas in the deep BEC regime. In this
limit, all fermions are paired into composite bosons (apart from
exponentially rare fluctuations). Inter-boson interactions occur
only via rare collisions. In the diagrammatic language, the
existence of compact composite bosons is reflected by the fact
that the dominant diagrams are the multi-ladder ones (see
\fref{fig:braids}A), with the typical $\mathbf{x}$-span of each
ladder being of the order of $L$, and the typical $\tau$-span of
the order of $\beta$. Each ladder of such a diagram in fact mimics
a world line of a composite boson.

The analogy between ladder parts of multi-ladder diagrams and
world lines of composite bosons helps elucidating the structure of
typical diagrams at resonance. Indeed, at the two-particle
resonance composite bosons exist only virtually, but at the same
time the energy cost of creating a boson is zero. Thus,
multi-ladder diagrams still dominate, but the $\mathbf{x}$- and
$\tau$-span of individual ladders becomes finite and related to
the inter-particle distance, see \fref{fig:braids}B. Resonant
ladders also become more ``dilute'' since the average distance
between vertices diverges, but they still contain many vertices
since the resonance formation involves distances much smaller than
$\nu^{-1/3}$.

Proposing diagrams which disregard the nature of the resonant
interaction results in small acceptance ratios.
To significantly reduce this type of slowing down, we resort to a
worm-type updating scheme which allows one to, loosely speaking,
directly ``draw'' the multi-ladder diagrams.

\subsubsection{Worm creation/annihilation updates}

These updates are the necessary ingredient of the worm-type
scheme, since they switch between the diagonal (\ref{conf-p}) and
off-diagonal (\ref{conf-p-corr}) diagrams.
These updates transform diagrams as follows:
\begin{eqnarray}
\fl
\mathfrak{S}^{(Z)} \ni \mathcal{S}_p = \{ \dots,\, (\mathbf{x}_1,\tau_1),\, \dots \}%
\leftrightarrows  \\
\{ \dots,\, (\mathbf{x}_1,\tau_1),\, %
P(\mathbf{x},\tau),\, P^{\dagger}(\mathbf{x}',\tau'),\, \dots \} = %
\widetilde{\mathcal{S}}_p \in \mathfrak{S}^{(G)}\; .  %
\label{cre-ann}
\end{eqnarray}

To create a pair of two-point vertices $P(\mathbf{x},\tau)$
and $P^\dagger(\mathbf{x}',\tau')$, one first selects $\mathbf{x}$
and $\tau$ uniformly over the spatial lattice and $(0,\beta)$
interval, respectively. One then selects $\mathbf{x}'$ among the
lattice sites in the spatial cube with the edge length $l$
centered around $\mathbf{x}$ with equal probabilities, and selects
$\tau$ uniformly on the interval $(\tau-\Delta \tau/2,
\tau+\Delta\tau/2)$. The values of $l$ and $\Delta\tau$ are
arbitrary. The overall probability density is thus given by:
\begin{equation}
W(\mathcal{S}_p \to \widetilde{\mathcal{S}}_p )\; =\;
\frac{1}{\beta L^d} \, \frac{1}{\Delta\tau l^d}\; .
\label{prob-ca}
\end{equation}
The reverse update attempts to remove $P(\mathbf{x},\tau)$ and
$P(\mathbf{x}',\tau')$ provided $|\mathbf{x}_\alpha - \mathbf{x}'_\alpha| <l/2$
($\alpha = 1,2,3$)
and $|\tau-\tau'| < \Delta\tau/2$, as prescribed by the balance
requirement.

The solution to the detailed balance equation (\ref{balance-eq})
then reads:
\begin{equation}
\mathcal{R} = \left| \frac{\det
\mathbf{A}(\widetilde{\mathcal{S}}_{p})}{ \det
\mathbf{A}(\mathcal{S}_{p}) }  \right|^2  \beta L^d \Delta\tau l^d
\zeta\; .
\label{R-ca}
\end{equation}

In order to maximize the efficiency of the numerical scheme, it is
advantageous to have the acceptance probabilities
(\ref{Metropolis_1})-(\ref{Metropolis_2}) to be of the order
unity. On the other hand, the explicit macroscopically large
factors in the right-hand side of Eq.\ (\ref{R-ca}) render the
acceptance probabilities macroscopic. The use of the extra
weighting factor $\zeta$ is now clear: by choosing $1/\zeta=\beta
L^d \Delta\tau l^d /\widetilde{\zeta}$ one removes undesired
factors from the right-hand side of Eq.\ (\ref{R-ca}). The
remaining freedom of choosing $\widetilde{\zeta}$ has to be
employed to further fine-tune the acceptance probabilities. The
required values of $\widetilde{\zeta}$ are typically of the order
of unity.

\subsubsection{Adding/removing vertices within the worm framework}
\label{subsec:leaps}

These updates are central for the method since they change the
diagram order. The main idea of these updates is: given a pair of
``worm'' two-point vertices $P$ and $P^\dagger$, use, say,
$P^\dagger(\mathbf{x},\tau)$ as a tip of a magic pen to add or
remove vertices from the diagram (note that it suffices to use
either $P$ or $P^\dagger$ as a dynamical variable, the choice
being only a matter of taste):
%
\begin{eqnarray}
\fl \widetilde{\mathcal{S}}_p =
\{\dots,\, (\mathbf{x}_1,\tau_1),\, P^\dagger(\mathbf{x},\tau),\,\dots \} %
\leftrightarrows\\  %
\{\dots,\, (\mathbf{x}_1,\tau_1),\, (\mathbf{x},\tau),\,
P^\dagger(\mathbf{x}_\mathrm{new},\tau_\mathrm{new}),\,\dots \} =
\widetilde{\mathcal{S}}'_{p+1}\; . %
\label{leaps}
\end{eqnarray}
 We have employed two versions of these updates.

\paragraph{High-density version}

In the most simple yet useful version of a forward update, one
selects $\tau_\mathrm{new}$ and $\mathbf{x}_\mathrm{new}$
uniformly in the interval $(\tau-\Delta \tau/2, \tau + \Delta
\tau/2)$, and the spatial cube of the edge length $l$ centered at
$\mathbf{x}$, correspondingly. In the reverse update one selects a
vertex to be removed at random out of $m$ choices, where $m$ is
the number of vertices $(\mathbf{x}_i,\tau_i)$ such that
$|(\mathbf{x}_i)_\alpha - (\mathbf{x}_\mathrm{new})_\alpha | < l/2$ and $| \tau_i -
\tau_\mathrm{new} | < \Delta \tau/2 $. The solution to the
detailed balance equation then reads
\begin{equation}
\mathcal{R}\;  =\;  \left| \frac{\det
\mathbf{A}(\widetilde{\mathcal{S}}_{p+1}')} { \det
\mathbf{A}(\widetilde{\mathcal{S}}_{p}) } \right|^2 \,
\frac{(-U)\Delta\tau l^d}{m}\; .
\label{R-le}
\end{equation}

\paragraph{Low-density version}

The strategy just described works well close to half-filling. In
the low-density regime the multi-ladder diagrams dominate and the
above scheme is not optimal for a number of reasons: (i) it does
not respect the diagram structure which should feature (at least
locally) well-defined vertex chains; (ii) the probability density
according to which the value of $\tau_\mathrm{new}$ is proposed
should favour shifting $P^\dagger$ towards smaller $\tau$-s. To
see this, consider two particles in vacuum: the matrix elements of
$P^\dagger(\mathbf{x},\tau_1) P(\mathbf{x},\tau_2)$ simply equal
zero for $\tau_2 < \tau_1$.]; (iii) both large and small values of
$\delta\tau = \tau - \tau_\mathrm{new}$ have to be accounted for.
Consider two fermions in vacuum again. The local probability
density for a shift of length $\delta\tau$ and
$\mathbf{x}_\mathrm{new}=\mathbf{x}$ is proportional to the free
diffusion propagator: $w(\delta \tau) \propto \delta\tau^{-3/2}$,
which is singular as $\delta\tau \to 0$. However, the mean value
of $\delta\tau$ for this distribution, $\int
 w(\delta\tau) \delta\tau \, \rmd(\delta\tau)$,  diverges at the upper limit.
These properties of resonant ladders are ignored when shifts are proposed
uniformly over some $\pm \Delta\tau$ interval.

The requirements (ii) and (iii) are best met if, when adding a
vertex, a new position of $P^\dagger$ is proposed according to the
probability density [cf. Eq.\ (\ref{leaps})]

\begin{equation}
w(\mathbf{x}_\mathrm{new}-\mathbf{x};\delta \tau) \propto %
\left\{%
\begin{array}{ll}
\left| G^{(0)}(\mathbf{x}-\mathbf{x}_\mathrm{new},\,  \tau -%
\tau_\mathrm{new}) \right|^2, & \mathrm{if~~} \tau_\mathrm{new}>\tau\;,\\
0, & \mathrm{otherwise.}
\end{array}%
\right.
\label{W-leap}
\end{equation}
Since the free single-particle propagator is known only
numerically, we use Eq.\ (\ref{W-leap}) on a uniform mesh with
some step $\sigma$: $\delta\tau_j = \sigma j$.  That is, we
tabulate the values $w_j = w(\mathbf{y}; \delta\tau_j)$ prior to
the start of computations; we then select $\mathbf{y}$ and
$\delta\tau_j$ according to the (normalized) distribution
$w(\mathbf{y};\delta\tau_j)$. The proposed new position of
$P^\dagger$ is then chosen as $\mathbf{x}_\mathrm{new} =
\mathbf{x}+\mathbf{y}$, and $\tau_\mathrm{new}=\tau+\delta\tau_j +
\Delta$, where $\Delta$ is uniform over the interval
$[-\sigma/2,\sigma/2]$. We find that $\sigma \sim 1/5U$ produces
good results.

In order to meet the requirement (i), the notion of the
closest---in terms of a certain distance---neighboring vertex is
introduced: the reverse update $\mathcal{S}'_{p+1} \to
\mathcal{S}_p$ \textit{deterministically} deletes the closest
neighbor of $P^\dagger$. The detailed balance than requires that
the forward update should be automatically rejected whenever
$(\mathbf{x},\tau)$ is {\it not} the closest neighbor of
$P^\dagger(\mathbf{x}_\mathrm{new},\tau_\mathrm{new})$, cf. Eq.\
(\ref{leaps}). We define the distance between vertices using a
simple Euclidean norm $\|(\mathbf{x},\tau) -
(\mathbf{x}_\mathrm{new},\tau_\mathrm{new}) \| =
(\mathbf{x}-\mathbf{x}_\mathrm{new})^2/L^2 +
(\tau-\tau_\mathrm{new})^2/\beta^2$.

The acceptance ratio function (\ref{balance-eq}) is then
\begin{equation}
\mathcal{R} = \left| \frac{\det \mathbf{A}(\widetilde{\mathcal{S}}_{p+1}')}{
\det \mathbf{A}(\widetilde{\mathcal{S}}_{p}) }  \right|^2
\frac{(-U)\sigma}{w_j}\; ,
\label{R-leap}
\end{equation}
where $w_j$ is given by Eq.\ (\ref{W-leap}).

The  scheme (\ref{R-leap}) in the dilute regime typically yields a
an order of magnitude gain in efficiency, as compared to
the scheme (\ref{R-le}).

\subsubsection{Shifting the worm two-point vertices}

Clearly, the scheme (\ref{R-leap}) should be supplemented by an
update which allows for an easy change of the closest neighbor.
The following update performs the task:
\begin{equation}
\widetilde{\mathcal{S}}_p \; =\;  \{\dots,\, P(\mathbf{x},\tau),\,\dots \}
\leftrightarrows %
\{\dots,\, P(\mathbf{x}',\tau'),\,\dots \}\;  =\;
\widetilde{\mathcal{S}}'_p\; .
\label{hop}
\end{equation}
We select $\mathbf{x}'$ among the nearest neighbors of the lattice
site $\mathbf{x}$ with equal probabilities and $\tau'$ uniformly
in some interval around $\tau$.

This update is self-balanced, and the acceptance ratio function
$\mathcal{R}$ is simply
\begin{equation}
\mathcal{R}\;  =\;  \left| \frac{\det
\mathbf{A}(\widetilde{\mathcal{S}}_{p})}{ \det
\mathbf{A}(\widetilde{\mathcal{S}}'_p)}
\right|^2\; .
\label{R-hop}
\end{equation}



\end{document}